\documentclass[11pt, letterpaper]{article}
\usepackage{graphicx} 

\usepackage[english]{babel}
\usepackage[utf8]{inputenc}
\usepackage[T1]{fontenc}
\usepackage{lmodern}
\usepackage[letterpaper, margin=1in]{geometry}
\usepackage{graphicx}
\usepackage{framed}
\usepackage[framemethod=tikz]{mdframed}
\usepackage{todonotes}
\usepackage{color}
\usepackage{microtype}

\definecolor{darkgreen}{rgb}{0,0.5,0}
\definecolor{darkgray}{rgb}{0.2,0.2,0.2}
\usepackage{hyperref}
\hypersetup{
    unicode=false,          
    colorlinks=true,        
    linkcolor=blue,         
    citecolor=purple,       
    filecolor=magenta,      
    urlcolor=cyan           
}

\usepackage{amsthm}
\usepackage{amsmath}
\usepackage{amssymb}
\usepackage{amsfonts}
\usepackage{mathrsfs}
\usepackage{mathtools}
\usepackage{verbatim}
\usepackage{footnote}

\usepackage{thmtools}

\usepackage{algorithm}
\usepackage{algorithmicx}
\usepackage[noend]{algpseudocode}
  
\usepackage{lineno}
\usepackage{caption}
\usepackage{enumerate}

\usepackage{tikzsymbols}
\usepackage{nicefrac}
\usepackage[most]{tcolorbox}

\usepackage{subcaption}
\usepackage{pdfpages}
\usepackage{makecell}

\newtheorem{theorem}{Theorem}[section]
\newtheorem{lemma}[theorem]{Lemma}
\newtheorem{corollary}[theorem]{Corollary}
\newtheorem{definition}[theorem]{Definition}

\newtheorem{claim}[theorem]{Claim}

\newtheorem*{remark*}{Remark}

\usepackage[capitalize, nameinlink]{cleveref}
\usepackage[section]{placeins}
\crefname{theorem}{Theorem}{Theorems}
\Crefname{lemma}{Lemma}{Lemmas}
\Crefname{invariant}{Invariant}{Invariants}
\Crefname{claim}{Claim}{Claims}
\Crefname{observation}{Observation}{Observations}
\Crefname{algorithm}{Algorithm}{Algorithms}
\Crefname{figure}{Figure}{Figures}
\Crefname{challenge}{Challenge}{Challenges}

\DeclareMathOperator{\poly}{poly}

\newcommand{\eps}{\epsilon}
\newcommand{\alp}{\alpha}

\newcommand{\defeq}{\overset{\text{\tiny def}}{=}}
\newcommand{\inhib}{\mathsf{inhib}}
\newcommand{\rank}{\mathsf{rank}}

\newcommand{\E}{\mathbb{E}}

\newcommand{\RGMIS}{\mathsf{RGMIS}}

\newcommand{\calE}{\mathcal{E}}
\newcommand{\calF}{\mathcal{F}}
\newcommand{\calG}{\mathcal{G}}

\newcommand{\calA}{\mathcal{A}}
\newcommand{\paren}[1]{\left( #1 \right)}

\newcommand{\ceil}[1]{\lceil #1 \rceil}
\newcommand{\Galive}{G_{\textnormal{alive}}}

\newcounter{resultctr}

\newtcolorbox{resultbox}[1][]{%
  enhanced,
  colback=gray!10,
  colframe=gray!60,
  coltext=black,
  boxrule=0.6pt,
  arc=2mm,
  left=1.2mm,
  right=1.2mm,
  top=1.0mm,
  bottom=1.0mm,
  before title={\refstepcounter{resultctr}}, 
  title={Main result\if\relax\detokenize{#1}\relax\else:~(#1)\fi},
  fonttitle=\bfseries,
  coltitle=black,
  attach title to upper,
  after title=\par\smallskip,
}

\title{Tight Query Depth of Recursive \\
Randomized Greedy MIS}
\author{ Slobodan Mitrović\thanks{Supported by the Google Research Scholar and NSF Faculty Early Career Development Program No.~2340048. e-mail: \texttt{smitrovic@ucdavis.edu}} \\ UC Davis 
\and Wen-Horng Sheu\thanks{Supported by the NSF Faculty Early Career Development Program No.~2340048. e-mail: \texttt{wsheu@ucdavis.edu}} \\ UC Davis}
\date{}

\begin{document}

\maketitle

\begin{abstract}
The randomized greedy maximal independent set (RGMIS) algorithm processes
vertices in a uniformly random order and greedily constructs an MIS.
Its recursive formulation decides whether a vertex belongs to this MIS
by querying lower-ranked neighbors in increasing rank order, stopping
when a neighbor in the MIS is found. 
We prove that, on every $n$-vertex
graph, this recursive algorithm has $O(\log n)$ query depth with high
probability, even without memoization. 
The bound holds simultaneously for all starting vertices and is asymptotically tight even on bipartite graphs.
The proof combines a bound on the total expected number of query paths with an exposure argument: paths with many two-edge shortcuts are unlikely, while the remaining paths admit sufficiently many applications of a two-hop probability bound.
\end{abstract}

\clearpage

\tableofcontents

\clearpage

\section{Introduction}

A maximal independent set (MIS) of a graph is an inclusion-wise maximal set of pairwise nonadjacent vertices. 
Computing an MIS is a central symmetry-breaking problem in graph algorithms, with fundamental roles in parallel, distributed, and sublinear computation~\cite{karp1985fast,luby1985simple,alon1986fast,calkin1990probabilistic,panconesi1992improved,naor1993can,kuhn2005fast,blelloch2012greedy,levi2015local,ghaffari2016improved,fischer2019tight,ghaffari2022local,khoury2026breaking}.

Among the simplest and most extensively studied approaches to computing an MIS is the randomized greedy MIS (RGMIS) algorithm. 
The algorithm samples a uniformly random permutation $\pi$ of the vertices and processes them in this order, adding a vertex if none of its previously processed neighbors has already been selected. 
Although sequential in its description, RGMIS admits a natural parallel execution. 
The \emph{parallel depth} of this execution has been studied since the 1980s, first on random graphs~\cite{coppersmith1989parallel,calkin1990probabilistic,calkin1992expected}, and later on arbitrary graphs under a uniformly random permutation. 
Blelloch, Fineman, and Shun proved an $O(\log^2 n)$ high-probability bound on the parallel depth~\cite{blelloch2012greedy}, while Fischer and Noever subsequently established the tight $\Theta(\log n)$ bound~\cite{fischer2019tight}.

RGMIS also admits a natural recursive formulation: to determine whether a vertex $v$ belongs to the resulting MIS, one recursively queries its lower-ranked neighbors in increasing rank order, stopping once a neighbor in the MIS is found. 
Since its introduction in sublinear-time graph algorithms~\cite{nguyen2008constant,yoshida2009improved}, this recursive process and closely related variants have become useful tools for vertex cover and matching~\cite{onak2012near,behnezhad2021time}, correlation clustering~\cite{behnezhad2022almost}, stochastic matching~\cite{azarmehr2025stochastic}, and metric Steiner problems~\cite{mahabadi2025sublinear,mahabadi2026sublinear}. 
The natural complexity measure of this formulation is its \emph{query depth}, namely, the maximum recursion depth over all queried vertices. 
Parallel depth does not directly imply query depth for RGMIS: Behnezhad constructed an instance whose parallel depth is $1$ but query depth is $\Omega(n)$ for \textbf{the same} vertex ordering~\cite{behnezhad2021time}.\footnote{In contrast, Behnezhad proved that for randomized greedy maximal matching, every query path has length at most $2\rho+1$, where $\rho$ is the parallel depth under the same edge ordering~\cite{behnezhad2021time}.}
The best previously known general upper bound on the query depth of RGMIS is $O(\log^2 n)$ with high probability~\cite{blelloch2012greedy,behnezhad2021massively}, while the parallel depth is known to be quadratically smaller~\cite{fischer2019tight}. 
This raises the following natural question:
\begin{center}
	\emph{Is this quadratic gap inherent? Does RGMIS have $O(\log n)$ query depth with high probability?}
\end{center}

\subsection{Our result}

\begin{table}[h]
\centering
\setlength{\extrarowheight}{4pt}
\begin{tabular}{|c|c|c|c|c|}
\hline
Reference & Graph class & Depth type & Result \\ \hline \hline
\cite{blelloch2012greedy} & General & Parallel & $O(\log \Delta \cdot \log n)$ \\ \hline
\cite{fischer2019tight} & General & Parallel & $\Theta(\log n)$ \\ \hline
\cite{blelloch2012greedy} + \cite{behnezhad2021massively} & General & Query & $O(\log^2 n)$ \\ \hline
\cite{behnezhad2021time} & Line graphs & Query & $O(\log n)$ \\ \hline
\cref{thm:main} & General & Query & $\Theta(\log n)$ \\ \hline
\end{tabular}
\caption{\label{tab:result}A summary of known results.}
\end{table}

\cref{tab:result} outlines prior work and our result.
As our main result, we show that RGMIS has $O(\log n)$ query depth. 
\begin{resultbox}[Informal version of \cref{thm:main}]
    With high probability, RGMIS has $O(\log n)$ query depth.
\end{resultbox}
As we show in \cref{sec:lower-bound}, this upper bound is asymptotically tight. 
The previously best-known upper bound on the query depth is $O(\log^2 n)$, which follows from \cite{blelloch2012greedy}.
See, e.g., \cite[Section 5]{behnezhad2021massively}.
We emphasize that this result does not assume that RGMIS recursive calls are cached/memoized.

The RGMIS is a natural random process studied in a long sequence of work and has been used as a subroutine in several algorithms. The studies on its average query tree size \cite{yoshida2009improved} and parallel depth \cite{fischer2019tight} have proved influential.
However, our understanding of the recursive query process is still quite limited, especially in the high-probability regime.
Our tight analysis fully characterizes the query depth of RGMIS in this regime.
Upper bounds on the query depth have been used in \cite{behnezhad2021time,behnezhad2021massively}, and we hope that our new result will inspire more applications in the field.

\subsection{Overview of our approach}

\subsubsection{RGMIS on triangle-free graphs}
\label{sec:overview-tri-free}




Our analysis is much easier to understand in triangle-free graphs. In the following, we outline the key techniques used in the analysis and highlight where the triangle-free assumption is used. In \cref{sec:overview-general}, we outline how to remove the assumption.

Fix an increasing sequence of \emph{positions} $1 \le p_1< \cdots < p_L \le n$ and bound the probability that the vertices occupying these positions form a query path $P=(\pi(p_1),\pi(p_2),\ldots,\pi(p_L))$.
We analyze the uniformly random permutation $\pi$ through an equivalent deferred-decisions process, in which positions -- not necessarily consecutive ones -- are exposed one by one. 
Exposing a position $p$ means revealing the previously unknown vertex $\pi(p)$ occupying that position. 
Conditioned on all positions exposed so far, the unexposed positions still contain a uniformly random permutation of the remaining vertices.

\paragraph{Survival conditions.}
For the prescribed positions $p_1,\ldots,p_L$ to yield a query path, the vertices revealed at consecutive path positions must be adjacent \textbf{and} must satisfy the appropriate survival conditions. 
For example, if $\pi(p_i)$ queries $\pi(p_{i-1})$, then no neighbor of $\pi(p_i)$ appearing before position $p_{i-1}$ can belong to the MIS. 
To capture this requirement, we associate with each path position a set of earlier ``protected'' positions. 
We expose positions $p_1, \ldots, p_L$ one by one.
After the vertex at the current path position has been exposed, we expose its protected positions and require that this vertex remain alive throughout these exposures. 
Here, a vertex is alive at time $t$ if none of its neighbors appearing before position $t$ belongs to the MIS.

\paragraph{One-hop lemma (\cref{lem:one-hop}).}
Thus, the basic local event needed to extend the candidate path by one edge is the following: (i) the vertex $\pi(p_i)$ already exposed at the current path position remains alive while its protected positions are being exposed, and (ii) the next path position then exposes an alive neighbor of $\pi(p_i)$. 
We bound the probability of this event using our one-hop lemma.

Suppose that $\pi(p_i)$'s protected set contains $\ell$ positions and that $\pi(p_i)$ has roughly $d$ alive neighbors. 
For this vertex to remain alive while the protected positions are exposed, none of these neighbors should be selected, which occurs with probability of roughly $\exp(-d\ell/n)$. 
After these exposures, $\pi(p_{i+1})$ should be a neighbor of $\pi(p_i)$, which occurs with a probability of roughly $d/n$. 
The product is maximized when $d\approx n/\ell$, yielding a bound of order $1/(e\ell)$, independent of the maximum degree. 
The formal lemma establishes this bound even though the set of alive neighbors changes during the exposure process.


The one-hop bound has essentially the right scale to balance the number of possible choices of the path positions, but by itself it provides no slack for a high-probability bound. 
To see this, let $L$ be the length of the candidate query path, and let the protected set sizes be roughly $\ell \approx n/L$. 
For any fixed choice of the $L$ path positions, multiplying the one-hop bounds gives approximately
\[
	\left(\frac{1}{e\ell}\right)^L \approx \left(\frac{L}{en}\right)^L.
\]
On the other hand, the number of possible choices of these positions is at most
\[
	\binom{n}{L} \le \left(\frac{en}{L}\right)^L.
\]
Thus, after taking a union bound, these two factors essentially cancel. In particular, the one-hop lemma alone does not yield an inverse-polynomial failure probability.

\paragraph{Two-hop lemma (\cref{lem:two-hop}).}
We therefore need to gain an additional constant factor for a linear number of hops. 
This is where the triangle-free assumption becomes important:
In a triangle-free graph, every three consecutive vertices in a query path form an induced path.

To see why this helps, let $v=\pi(p_i)$ and consider its alive neighbors, which are the possible choices for $\pi(p_{i+1})$. 
As discussed above, the one-hop bound is close to its worst-case value only when $v$ has roughly $n/\ell$ alive neighbors. 
After $\pi(p_{i+1})=u$ is exposed, the second one-hop bound can be close to its worst-case value only if $u$ also has roughly $n/\ell$ suitable alive neighbors. 
However, since the three vertices must form an induced path, $\pi(p_{i+2})$ cannot be adjacent to $v$; it must be an alive neighbor of $u$ at distance exactly two from $v$.
This separation between the one- and two-hop neighborhoods is important. 
A vertex at distance two from $v$ may eliminate $u$ without eliminating $v$, so such eliminations remain possible even after conditioning on $v$ surviving the exposures of its protected set.

\begin{figure}[h]
\centering
\begin{tikzpicture}[x=1cm,y=1cm,
    every node/.style={font=\small},
    vertex/.style={circle,draw,minimum size=5.5mm,inner sep=0pt},
    cand/.style={circle,draw,fill=blue!15,minimum size=5.5mm,inner sep=0pt},
    bad/.style={circle,draw,fill=red!15,minimum size=5.5mm,inner sep=0pt},
    helper/.style={circle,draw,fill=gray!15,minimum size=5.5mm,inner sep=0pt}
]

\node at (3.2,2.2) {\textbf{Few candidates}};

\node[vertex] (vL) at (0,0) {$v$};

\node[vertex] (u1L) at (2,1.2) {$u_1$};
\node[cand]   (u2L) at (2,0.4) {$u_2$};
\node[vertex] (u3L) at (2,-0.4) {$u_3$};
\node[vertex] (u4L) at (2,-1.2) {$u_4$};

\node[helper] (w1L) at (4,1.0) {$w_1$};
\node[helper] (w2L) at (4,0.0) {$w_2$};
\node[helper] (w3L) at (4,-1.0) {$w_3$};

\draw (vL) -- (u1L);
\draw (vL) -- (u2L);
\draw (vL) -- (u3L);
\draw (vL) -- (u4L);

\draw (u2L) -- (w1L);
\draw (u2L) -- (w2L);
\draw (u2L) -- (w3L);

\node[align=center] at (2,-2.5) {$\pi(p_{i+1})$ must land on one of\\the few blue candidate vertices};

\draw[dashed] (5.3,-2.4) -- (5.3,2.4);

\node at (8.8,2.2) {\textbf{Many candidates}};

\node[vertex] (vR) at (6.2,0) {$v$};

\node[cand] (u1R) at (8.2,1.4) {$u_1$};
\node[cand] (u2R) at (8.2,0.5) {$u_2$};
\node[cand] (u3R) at (8.2,-0.5) {$u_3$};
\node[cand] (u4R) at (8.2,-1.4) {$u_4$};

\node[helper] (w1R) at (10.3,1.5) {$w_1$};
\node[bad]    (w2R) at (10.3,0.5) {$w_2$};
\node[helper] (w3R) at (10.3,-0.5) {$w_3$};
\node[helper] (w4R) at (10.3,-1.5) {$w_4$};

\draw (vR) -- (u1R);
\draw (vR) -- (u2R);
\draw (vR) -- (u3R);
\draw (vR) -- (u4R);

\draw (u1R) -- (w1R);
\draw (u1R) -- (w2R);

\draw (u2R) -- (w1R);
\draw (u2R) -- (w2R);
\draw (u2R) -- (w3R);

\draw (u3R) -- (w2R);
\draw (u3R) -- (w3R);
\draw (u3R) -- (w4R);

\draw (u4R) -- (w3R);
\draw (u4R) -- (w4R);

\draw[very thick,red,->] (11.0,0.7) -- (w2R);
\node[align=left,red] at (12.7,0.9) {\scriptsize exposed vertex\\[-1mm]\scriptsize eliminating candidates};

\node[align=center] at (8.9,-2.5) {distance-two vertices can eliminate many\\candidates without eliminating $v$};

\end{tikzpicture}
\caption{A schematic view of the two cases in the two-hop analysis. 
Left: if only a few neighbors of $v$ have many alive neighbors at distance two from $v$, then $\pi(p_{i+1})$ is unlikely to be one of them. 
\\
Right: if many such neighbors exist, then exposing vertices at distance two from $v$ can eliminate many candidates while leaving $v$ alive, since these distance-two vertices are not adjacent to $v$.}
\label{fig:two-hop-cases}
\end{figure}
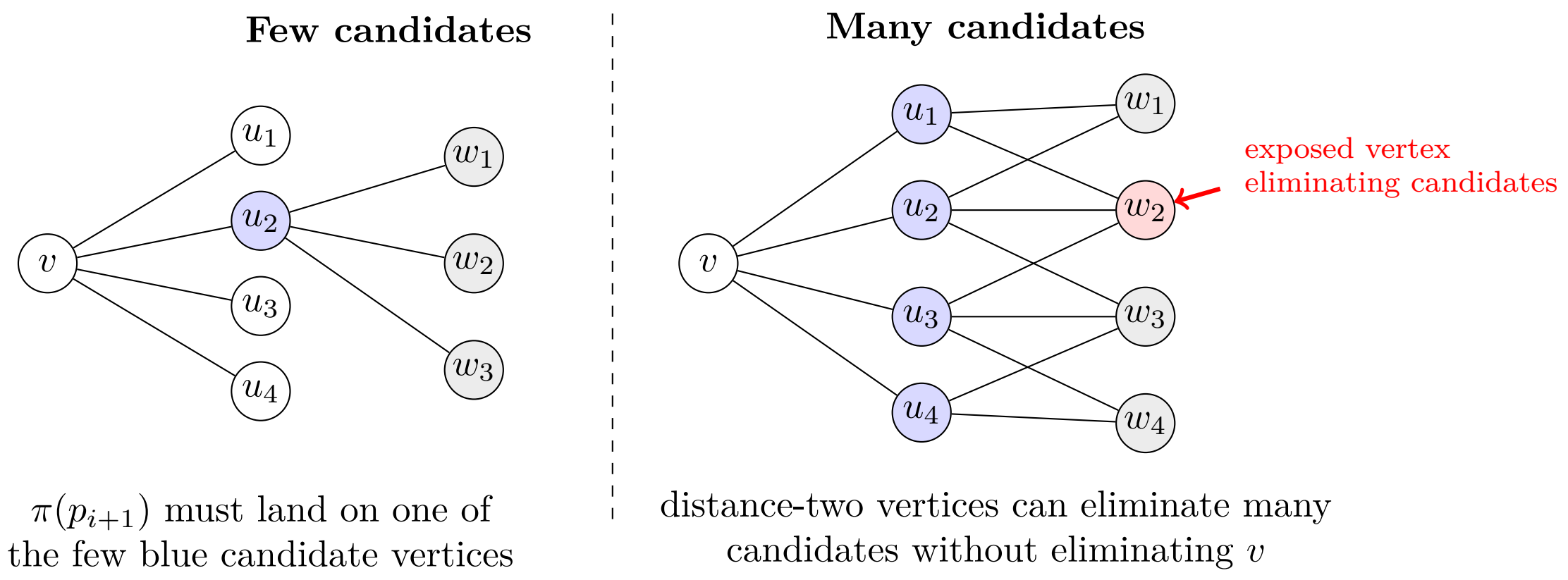

Thus, only those neighbors $u$ of $v$ having many alive neighbors at distance two from $v$ can support a near-worst-case second hop. 
\begin{itemize}
	\item If there are few such vertices $u$, then $\pi(p_{i+1})$ is unlikely to be one of them, and the second hop loses a constant factor. 
	\item If there are many such vertices $u$, then the exposures used to keep $v$ alive have many opportunities to eliminate vertices like $u$. 
	Consequently, with constant probability, ``too few'' candidates remain when $\pi(p_{i+1})$ is exposed, again yielding a constant-factor improvement.
\end{itemize}
Hence, the two hops cannot simultaneously attain their individual worst-case bounds. This behavior is illustrated in \cref{fig:two-hop-cases}.

Our two-hop lemma formalizes this tradeoff: when the two protected sets have the same size $\ell$, the probability of the joint event is at most
\[
	\frac{1-\epsilon}{(e\ell)^2}
\]
for some constant $\epsilon>0$. 
Applying this bound to a linear number of pairs produces the additional factor $(1-\epsilon)^{\Omega(L)}$ needed for the final union bound.


\paragraph{Summary.} 
Hence, at a high level, the one-hop lemma pays for the combinatorial freedom in choosing the path positions: after the union bound, it leaves us roughly at break-even. 
The role of the two-hop lemma is to turn this break-even estimate into a concrete probabilistic decay. 
Namely, it leverages the fact that two consecutive candidate edges cannot simultaneously exhibit their worst-case behavior, so each such pair improves the probability bound by a constant factor.
These constant-factor improvements accumulate along the path, producing the exponentially small factor needed for the high-probability bound.
\cref{sec:interval-division} organizes the position exposures so that these two estimates can be applied repeatedly with appropriately sized protected sets.

\subsubsection{RGMIS in general graphs}
\label{sec:overview-general}

In the analysis above, a key property is that three consecutive vertices on a query path do not induce triangles, e.g., the one-hop neighborhood $N_1(v)$ does not contain any edges inside.
To extend this argument to general graphs, the main difficulty is that $N_1(v)$ may contain many edges. 
Thus, after exposing $\pi(p_{i+1})=u\in N_1(v)$, the next path vertex may still lie inside $N_1(v)$, forming a triangle with $v$ and $u$. 
This destroys the clean separation between the one- and two-hop neighborhoods used in the two-hop analysis. 

To overcome this difficulty, we show that most length-three sub-paths on a query path must be an induced path, assuming the query path is long enough.
This in turn implies that the two-hop lemma can be applied on most of the query path.
More precisely, one can fix a small subset $Q$ of length-three sub-paths that do not form induced paths, and apply the two-hop lemma when the current position does not intersect $Q$. The largest size of $Q$ will be small enough that it does not affect the probability bound by too much.

We next explain why the number of non-induced length-three sub-paths is bounded.
Suppose that $R = (v_1, v_2, \dots, v_L)$ is a query path and there is a set $X$ of \emph{shortcut edges} of the form $(v_i, v_{i+2})$, $i \in [1, L-2]$.
Observe that replacing the sub-path $(v_i, v_{i+1}, v_{i+2})$ with a single edge $(v_i, v_{i+2})$ gives a new query path.
Hence, if all sub-paths corresponding to $X$ are disjoint, replacing any subset of them gives a new query path.
These sub-paths may not be disjoint.
However, it is not hard to select $\lceil |X| / 2\rceil$ edges from $X$ such that their corresponding sub-paths are disjoint.
Therefore, the number of query paths is at least $2^{|X|/2}$.
On the other hand, it is known that the number of query paths is upper bounded by $\poly(n)$ with high probability;
this is a simple implication of the average complexity result by \cite{yoshida2009improved}.
Consequently, with high probability, no query path can have more than $\Omega(\log n)$ shortcut edges on it.
For sufficiently large $L = \Theta(\log n)$, the number of shortcut edges on any query path is only a small constant fraction of $L$.

\subsection{Related work}

\paragraph{Comparison to \cite{fischer2019tight}.}
The closest to our work is the approach by Fischer and Noever~\cite{fischer2019tight}.
The high-level perspective of our analysis for RGMIS  in triangle-free graphs follows their framework, although many technical changes are required to obtain the query depth guarantee even for triangle-free graphs.
The main difference arises in general graphs.
Namely, once three consecutive vertices on a query path induce a triangle, we need an approach that goes beyond Fischer and Noever: The clean separation between $N_1(v)$ and $N_2(v)$ is no longer available.

\paragraph{Other related work.}
Luby's celebrated parallel MIS algorithm can be viewed as the parallel greedy MIS process in which a fresh random ordering is sampled in every round, rather than fixing one random permutation throughout the execution~\cite{luby1985simple}. 
The process terminates in $O(\log n)$ rounds with high probability.
This algorithm, and the closely related independent work of Alon, Babai, and Itai~\cite{alon1986fast}, has been highly influential in parallel and distributed graph algorithms.

Nguyen and Onak introduced the recursive formulation of RGMIS in the context of constant-time approximation algorithms~\cite{nguyen2008constant}. 
They showed that, on graphs of maximum degree $\Delta$, deciding whether an arbitrary vertex belongs to the resulting MIS requires at most $2^{O(\Delta)}$ recursive calls in expectation. 
They further conjectured that examining the lower-ranked neighbors in increasing rank order and stopping after finding the first neighbor in the MIS yields a substantially better bound. 
Yoshida, Yamamoto, and Ito proved this conjecture in a strong average-case form: over a uniformly random permutation, the total number of recursive calls across all root vertices is at most $m$ in expectation, or equivalently at most $m/n$ for a uniformly random root vertex~\cite{yoshida2009improved}. 
They used this local oracle to obtain improved constant-time approximation algorithms for maximum matching and several other optimization problems on bounded-degree graphs.

Blelloch, Fineman, and Shun showed that the parallel depth of RGMIS is $O(\log^2 n)$ with high probability~\cite{blelloch2012greedy}; Fischer and Noever later improved this to the tight $\Theta(\log n)$ bound~\cite{fischer2019tight}. 
Behnezhad, Dhulipala, Esfandiari, {\L}{\k{a}}cki, Mirrokni, and Schudy observed that the analysis of \cite{blelloch2012greedy} can also be adapted to give an $O(\log^2 n)$ high-probability bound on the query depth~\cite[Section~5]{behnezhad2021massively}.

\section{Preliminaries}
\subsection{Notations}
Consider a simple undirected graph $G = (V, E)$ with $n\geq2$.
We write $[t]=\{1,\ldots,\lfloor t\rfloor\}$.
The length of a path is the number of vertices on it.
Let $n$ and $m$ denote, respectively, the number of vertices and edges of $G$.
Denote by $\Delta(H)$ the maximum degree of a graph $H$.
The maximum degree of $G$ is simply denoted by $\Delta$.

Let $\pi$ be a permutation of $V$ chosen uniformly at random.
All definitions in this section are defined with respect to $\pi$.
Denote by $\pi(i)$ the $i$-th vertex in $\pi$.
Let $\pi(l, r)$ be the set of vertices whose positions in $\pi$ are between $l$ and $r$.

Denote by $V^*$ the output of RGMIS.
This algorithm was already outlined in the introduction, and here we describe it more formally.
Let $S$ be an initially empty set.
Scan the vertices in the order of $\pi$.
When a vertex $v$ is scanned, we check if $S \cup \{v\}$ is an independent set, and if so, set $S \gets S \cup \{v\}$.

The above algorithm has several variants; its most basic recursive variant is given as \cref{alg:RGMIS-LCA}.
It determines whether a given vertex is in $V^*$ when the permutation is $\pi$.

\begin{algorithm}[H]
\begin{algorithmic}[1]
\medskip 
\Statex \textbf{Input:} a vertex $v \in V(G)$, a permutation $\pi$
\Statex \textbf{Output:} whether $v \in V^*$ when the permutation is $\pi$
\medskip
\Statex \hrule
\For {all neighbors $u$ of $v$ whose rank is smaller than $v$, in increasing order of their ranks}
    \If {$\mathsf{RGMIS}(u, \pi)$ is YES} \Comment{decide if $u$ is in RGMIS}
        \State \Return NO
    \EndIf
\EndFor
\State \Return YES
\end{algorithmic}
\caption{$\RGMIS:$ A recursive algorithm for computing RGMIS}
\label{alg:RGMIS-LCA}
\end{algorithm}

\begin{definition}[Rank, inhibitor]
    The rank of a vertex $v$, denoted by $\rank(v)$, is its position in $\pi$, that is, $\pi^{-1}(v)$.
    For a vertex $v \notin V^*$, the \emph{inhibitor} of $v$, denoted by $\inhib(v)$, is defined as the neighbor of $v$ in $V^*$ that has the smallest rank.
    For a vertex $v \in V^*$, we define the inhibitor of $v$ as $v$ itself.

    For a position $p \in [1, n]$, we define the \emph{inhibitor} of $p$ as the rank of the inhibitor of $\pi(p)$.
    The inhibitor of a position $p$ is denoted by $\inhib(p)$.
\end{definition}

\begin{definition}[Alive, dead]
    A vertex $v$ is called \emph{alive} at time $t$ if $\rank(\inhib(v)) \geq t$.
    In other words, $v$ is alive at $t$ if it is not removed after the greedy algorithm processes $\pi(1, t-1)$.
    A position $p$ is \emph{alive} at time $t$ if $\pi(p)$ is alive at time $t$.
    A vertex or position is called \emph{dead} if it is not alive.
\end{definition}

\begin{definition}[Y-position, N-position]
A \emph{Y-position} with respect to $\pi$ is a position $p \in [1, n]$ such that $\pi(p) \in V^*$.
An \emph{N-position} is a position $p$ such that $\pi(p) \notin V^*$.
\end{definition}

Consider a set of positions $P = \{p_1, p_2, \dots, p_L\}$ where $p_1 < p_2 < \dots < p_L$.
We define $\min P$ and $\max P$ as $p_1$ and $p_L$, respectively.
Denote $P(l, r) = \{p_l, p_{l+1}, \dots, p_r\}$ and $P(t) = p_t$.
For a set of positions $P'$, we say $P < P'$ if $\max P < \min P'$.
For a position $p$, we write $P < p$ if $\max P < p$.
Denote by $\pi(P)$ the set $\{\pi(p) \mid p \in P\}$.

Throughout, $\log$ denotes the base-two logarithm, and $\ln$ denotes the natural logarithm.


\subsection{Dependency paths and query paths}
\begin{definition}[Increasing path]
    An \emph{increasing path} with respect to $\pi$ is a path $(v_1, v_2, \dots, v_L)$ such that $\rank(v_1) < \rank(v_2) < \dots < \rank(v_L)$.
    We say a sequence of positions $p_1, p_2, \dots, p_L$ \emph{forms an increasing path} if $(\pi(p_1), \pi(p_2), \dots, \pi(p_L))$ is an increasing path.
\end{definition}

\begin{definition}[Query path]
    We say a sequence of positions $p_1, \dots, p_L$ \emph{forms a query path} with respect to $\pi$ if all of the following hold.
    \begin{itemize}
        \item $(p_1, p_2, \dots, p_L)$ forms an increasing path, and
        \item$p_{i-1} \leq \inhib(p_{i})$ holds for all $i > 1$.
    \end{itemize}
\end{definition}

Observe that a path $(\pi(p_L), \pi(p_{L-1}), \dots, \pi(p_1))$ is explored by the recursive RGMIS algorithm (\cref{alg:RGMIS-LCA}) if and only if $(p_1, p_2, \dots, p_L)$ forms a query path.

\begin{definition}[Query depth] 
The \emph{query depth} of $G$ with respect to $\pi$ is the maximum length of a query path.
For a subgraph $H$ of $G$, denote by $\sigma(H)$ the query depth of $H$ when the permutation on $V(H)$ is inherited from $\pi$.
Observe that the query depth equals the maximum recursion-stack size over all roots $v$ for the RGMIS oracle (\cref{alg:RGMIS-LCA}).
\end{definition}

A related notion is the \emph{dependency path}, which is a type of query path that alternates between Y-positions and N-positions; in addition, the predecessor of each N-position must be its inhibitor.

\begin{definition}[Dependency path, dependency depth]
    We say a sequence of positions $p_1, p_2, \dots, p_L$ forms a \emph{dependency path} if all of the following hold.
    \begin{itemize}
        \item $(p_1, p_2, \dots, p_L)$ forms a query path of odd length,
        \item $\pi(p_i) \in V^*$ if and only if $i$ is odd, and
        \item $p_{i-1} = \inhib(p_{i})$ for even $i$.
    \end{itemize}
    The \emph{dependency depth} with respect to $\pi$ is the maximum length of a dependency path.
\end{definition}

The dependency depth is an upper bound on the parallel depth, and it is known that the dependency depth is $O(\log n)$ with high probability, with a matching
worst-case lower bound~\cite{fischer2019tight}.

\begin{definition}[Induced path]
A sequence of positions $p_1, p_2, \dots, p_L$ is said to form an \emph{induced path} if both of the following hold:
\begin{itemize}
    \item $(\pi(p_i), \pi(p_{i+1})) \in E$ for $1 \leq i < L$, and
    \item $(\pi(p_i), \pi(p_{j})) \notin E$ for $j \neq i+1$, $1 \leq i \leq j \leq L$.
\end{itemize}
\end{definition}

\subsection{Known properties}

For positive integers $l\leq r\leq n$, define $\Galive(l,r)$ as the
subgraph induced by vertices in $\pi(l,r)$ alive at time $l$.
Define $G(l,r)$ as the subgraph induced by $\pi(l,r)$.
For real $l, r$ we denote by $\Galive(l,r)=\Galive(\lceil l\rceil,\lfloor r\rfloor)$,
and similarly for $G(l,r)$.
The graph $\Galive(t,n)$ is the \emph{alive graph at time $t$}.

The greedy process on $\Galive(l,r)$ in the inherited order agrees
with the full process on its vertices: vertices removed before time
$l$ cannot subsequently enter the MIS, and vertices ranked after $r$
cannot affect an earlier decision. In particular, inhibitors of
vertices in $\Galive(l,r)$ are the same as in the full graph.

\begin{lemma}[Degree reduction bound~\cite{blelloch2012greedy}]
\label{lem:degree-reduction}
For every integer $1\leq t<n$ and real $d>0$,
\[
 \Pr[\Delta(\Galive(t+1,n))\geq d]\leq n\exp(-td/n).
\]
Consequently, this degree is at most $8n\ln n/t$ with probability
at least $1-1/n^7$.
\end{lemma}

\begin{lemma}[Increasing-path bound~\cite{blelloch2012greedy}]
\label{lem:increasing-path}
Let $H$ have at most $n$ vertices and maximum degree at most
$D\log n$, for a fixed constant $D$. Under a uniformly random relative
order of its vertices, the longest increasing path has
$O(\log n)$ vertices with probability at least $1-1/n^7$.
\end{lemma}

For a permutation $\pi$ and vertex $v$, let $T(v,\pi)$ be the number
of query paths ending at $v$, including the singleton path $(v)$.
Equivalently, it is the total number of calls in the recursion tree
of $\RGMIS(v,\pi)$, including its root and counting repeated calls
with multiplicity. Write $Z(\pi)=\sum_{v\in V}T(v,\pi)$.

\begin{lemma}[\cite{yoshida2009improved}]\label{lem:YYI}
For a uniformly random permutation $\pi$ on $V$,
\[
 \E[Z(\pi)]\leq n+m.
\]
\end{lemma}

\begin{corollary}\label{cor:YYI}
For a uniformly random permutation $\pi$ on $V$,
\[
 \Pr\!\left[Z(\pi)>n^9\right]\leq \frac{1}{n^7}.
\]
\end{corollary}
\begin{proof}
By Markov's inequality,
\[
    \Pr[Z(\pi)>n^9]\leq\E[Z(\pi)]/n^9\leq1/n^7,
\]
where the first inequality is due to \cref{lem:YYI}, and the second inequality holds since $n + m \leq n^2$.
\end{proof}

The following lemma lets us split the permutation into two intervals.

\begin{lemma}[Query-depth division]\label{lem:query-depth-division}
For integers $1\leq l\leq t<r\leq n$,
\[
 \sigma(\Galive(l,r))
 \leq\sigma(\Galive(l,t))+\sigma(\Galive(t+1,r))+1.
\]
\end{lemma}
\begin{proof}
Let $P=(v_1,\ldots,v_L)$ be a query path in $\Galive(l,r)$, written
in increasing rank order. Let $s$ be the number of its vertices
of rank at most $t$, where $0 \leq s \leq L$.
Its first $s$ vertices form a query path in $\Galive(l,t)$ and hence $s \leq \sigma(\Galive(l, t))$.
For $i\geq s+2$, the definition of query path implies
\[
 \rank(\inhib(v_i))\geq\rank(v_{i-1})>t.
\]
Thus all vertices $v_{s+2},\ldots,v_L$ are alive at time $t+1$.
Observe that $(v_{s+2}, v_{s+3}, \dots, v_L)$ forms a query path in $\Galive(t+1, r)$, and hence its length is at most $\sigma(\Galive(t+1, r))$.
Only $v_{s+1}$, if it exists, remains unaccounted for, proving the inequality.
\end{proof}
\section{An upper bound on the query depth}
\label{sec:main}

In this section, we present our main result -- an $O(\log n)$ upper bound on query depth in general graphs.
More specifically, we prove the following theorem.

\begin{theorem}[Upper bound on query depth] \label{thm:main}
Let $G$ be a simple undirected graph on $n$ vertices, and let
$\pi$ be a uniformly random permutation of $V(G)$. There is an
absolute constant $C$ such that
\[
 \Pr_\pi[\sigma(G)>C\log n]\leq \frac{1}{n^5}.
\]
Thus the recursive $\RGMIS$ oracle has
$O(\log n)$ query depth simultaneously for all starting vertices,
with high probability.
\end{theorem}

\noindent
The theorem above is tight even on bipartite graphs, and is tight up to a $\Theta(\log \log n)$ factor on paths; see \cref{sec:lower-bound} for details.

\subsection{Formulation of the random process}

In the following, we outline the key techniques to prove \cref{thm:main}.
These techniques can be seen as extensions of those in \cite{blelloch2012greedy,fischer2019tight}.
Recall that $\pi$ is a permutation drawn uniformly at random.
The construction of $\pi$ can be equivalently rephrased as the following random process.
Initially, all $n$ entries of $\pi$ are \emph{unexposed} (or undetermined), and the ranks of all vertices are also unknown.
In each step, we \emph{expose} one position $p$, by fixing $\pi(p)$ as a vertex chosen uniformly at random from the set of vertices that are not in any exposed position.

We say a vertex is \emph{exposed} if its position in $\pi$ is already fixed.
This process is repeated on a fixed sequence of distinct positions $p_1, p_2, \dots, p_L$.
If $L = n$, the resulting permutation is a uniformly random permutation.
For two disjoint sets of positions $S, T$ and an event $X$, we use the notation
\[
\Pr_T[X \mid \pi(S)]
\]
to denote the probability of $X$ when exposing $\pi(T)$, conditioned on $\pi(S)$ having already been exposed.

Consider a fixed sequence of positions $P = (p_1, p_2, \dots, p_L)$, where $p_1 < p_2 < \dots < p_L$, and let $L = \lceil\sqrt{A} \cdot \log n\rceil$ for a sufficiently large constant $A$.
Our goal is to prove that the probability of having a query path of length $L$ is at most $1/\poly(n)$.
We make the following simplifying assumptions: $A \cdot \log n < p_1 < p_2 < \dots < p_L < \beta n$ for a sufficiently small constant $\beta \in (0, 1)$.
These assumptions can be removed by known reductions; see \cref{sec:assumptions} for details.
We also assume that, without loss of generality, $n \ge n_0$ for a sufficiently large constant $n_0$.


\subsection{Warm-up: an argument for bounded-degree graphs} \label{sec:warm-up}

When the maximum degree is bounded by $O(\log n)$, the following argument gives an $O(\log n)$ upper bound on the query depth.
This argument is not directly used in the rest of the proof, but is helpful in understanding the random process that we follow to prove our main claim.

Consider the process where we expose the positions $p_1, p_2, \dots, p_L$ in order.
To form a query path, a necessary condition is that there is an edge between $p_i$ and $p_{i+1}$ for all $i < L$.
Let $P^{(i)} = \{p_1, p_2, \dots, p_i\}$.
When a position $p_{i+1}$ is exposed, vertex $\pi(p_{i+1})$ is chosen uniformly at random from the $n - i$ vertices that are not in $\pi(P^{(i)})$.
Out of these vertices, at most $\Delta$ of them can be a neighbor of $\pi(p_i)$.
Hence, we have
\[
    \Pr_{p_{i+1}}[\text{there is an edge between $\pi(p_i)$ and $\pi(p_{i+1})$} \mid \pi(P^{(i)})] \leq \frac{\Delta}{n - i}, \text{ where }1 \leq i < L.
\]
Suppose that $L < n/2$. Then, the above probability is at most $2\Delta / n$.
Hence, $P$ forms a query path with probability at most $(\frac{2\Delta}{n})^{L-1}$.

The above argument holds for a single choice of $P$.
Note that there are $\binom{n}{L} \leq (\frac{en}{L})^L$ ways to choose $P$.
By union bounding over all possible choices, the probability of having a query path of length $L$ is at most $(\frac{en}{L})^L \cdot (\frac{2\Delta}{n})^{L-1} \leq (\frac{2e\Delta}{L})^L \cdot en$.
For $\Delta = O(\log n)$ and $L = \Theta(\log n)$, the probability is at most $1/n^5$, yielding the advertised high probability guarantee.

Note that this proof consists of two steps: (1) Upper bounding the probability that a fixed position sequence $P$ forms a query path, and (2) applying a union bound over all possible choices of $P$.
Our main proof follows these two steps, but is significantly more involved than the analysis of this toy example.
The analysis of these two steps is presented in \cref{sec:hop-lemmas,sec:interval-division}.

\subsection{The bounded-hop lemmas} \label{sec:hop-lemmas}

In the warm-up argument, the probability that $\pi(p_{i+1})$ is adjacent to $\pi(p_i)$ is at most $O(\Delta/n)$, which is insufficient when $\Delta=\Theta(n)$. 
Our refined argument, which we refer to the one-hop lemma, rests on the following key idea: in addition to $(\pi(p_i), \pi(p_{i+1}))$ being an edge, require that
\begin{center}
	$p_i$ is alive at time $p_{i-1}$.
\end{center}
The reason is that $\inhib(p_i) \ge p_{i-1}$ is necessary for $P$ to be a query path.

More formally, our argument additionally exploits the requirement that $\pi(p_i)$ has to \emph{remain alive} while a set of $\ell$ positions \emph{preceding} $p_{i-1}$ is exposed. 
Heuristically, if $\pi(p_i)$ has $d$ alive neighbors, surviving these $\ell$ exposures contributes a factor of $(1-d/n)^\ell \approx \exp(-d\ell/n)$, while subsequently exposing an alive neighbor contributes a factor roughly $d/n$. 
Their product is maximized when $d\approx n/\ell$, yielding a bound of order $1/(e\ell)$, independent of $\Delta$.
In the following lemma, those $\ell$ positions are contained in the set $P_1$.

\begin{lemma}[The one-hop lemma] \label{lem:one-hop}
For any $\beta\in(0,0.1)$, the following holds.
Consider a non-empty set of positions $P_1$ and two positions $p_1, p_2$ such that $P_1 < p_1 < p_2 < \beta n$.
Suppose that a set of positions $S \subseteq [\beta n]$ is already exposed, and that
\begin{itemize}
    \item[(A1)] $S$ contains $p_1$ and all positions in $[\max P_1] \setminus P_1$,
    \item[(A2)] $S$ does not contain any position in $P_1 \cup \{p_2\}$.
    \item[(A3)] For some integer $1\leq\tau<\min P_1$, we have $[\tau]\subseteq S$ and $\Delta(\Galive(\tau+1,n))\leq n/2$.
\end{itemize}
Let $\mathcal F$ be the event that $(\pi(p_1),\pi(p_2))\in E$
and both vertices are alive at time $\max P_1+1$. Then
\[
 \Pr_{P_1\cup\{p_2\}}[\mathcal F\mid\pi(S)]
 \leq\frac{1}{e\ell(1-\beta)},\qquad \ell=|P_1|.
\]
\end{lemma}

We remark that the positions in $P_1$ do not need to be contiguous.
In the lemma, (A3) is a simplifying assumption that each vertex has at most $n/2$ alive neighbors when $P_1 \cup \{p_2\}$ is exposed.
By \cref{lem:degree-reduction}, exposing a sufficiently long
$\Theta(\log n)$ prefix will guarantee that (A3) holds.
(See \cref{sec:interval-division} for details, where we set $\tau \approx 18 \ln n$.)

When applying the lemma to consecutive positions $p_i,p_{i+1}$ of the candidate query path, the auxiliary set $P_1$ will lie entirely before $p_{i-1}$. (See \cref{sec:interval-division}.)
The proof of \cref{lem:one-hop} is deferred to \cref{sec:one-hop}.

The one-hop bound alone is not strong enough for the final union bound: we need to gain an additional constant factor for a linear number of pairs of consecutive edges. 
Whenever a triple of consecutive query-path vertices has no shortcut
between its first and third vertices, that triple forms an induced path. 
Consequently, when two consecutive one-hop events corresponding to the edges $(p_i,p_{i+1})$ and $(p_{i+1},p_{i+2})$ are analyzed jointly, they cannot simultaneously attain the worst-case probability allowed by \cref{lem:one-hop}. 
The two-hop lemma quantifies this loss, yielding an additional factor of $1-\epsilon$ for each pair to which it is applied.
\begin{lemma}[The two-hop lemma] \label{lem:two-hop}
There exist constants $\beta \in (0, 0.1)$ and $\eps \in(0,0.1)$ such that the following holds.
Consider two sets of positions $P_1, P_2$ and three positions $p_1, p_2, p_3$ such that $P_1 < P_2 < p_1 < p_2 < p_3 < \beta n$, and $|P_1| = |P_2| = \ell$ for some $\ell > 10^3$.
Suppose that a set of positions $S \subseteq [\beta n]$ is already exposed, and that
\begin{itemize}
    \item $S$ contains $p_1$ and all positions in $[\max P_2] \setminus P_1 \setminus P_2$, and
    \item $S$ does not contain any position in $P_1 \cup P_2 \cup \{p_2, p_3\}$.
    \item For some integer $1\leq\tau<\min P_1$, we have $[\tau]\subseteq S$ and $\Delta(\Galive(\tau+1,n))\leq n/2$.
\end{itemize}
Define the event $\calG$ as the intersection of the following:
\begin{itemize}
    \item $(p_1, p_2, p_3)$ forms an induced path,
    \item $p_1$ is alive at time $\max P_1 + 1$, and
    \item $p_2$ and $p_3$ are alive at time $\max P_2 + 1$.
\end{itemize}
Then,
\[
    \Pr_{P_1 \cup P_2 \cup \{p_2, p_3\}}[\calG \mid \pi(S)] \leq \frac{1-\eps}{(e\ell)^2}.
\]
\end{lemma}

The proof of this lemma is deferred to \cref{sec:two-hop}.
We first show how the two lemmas are used to prove our main result.

\subsection{Probability upper bound via interval division} \label{sec:interval-division}

For a fixed permutation $\pi$ and a query path $P=(p_1,\ldots,p_L)$, a
\emph{shortcut} on $P$ is a position $p_i$, with $i\leq L-2$, such that
$(\pi(p_i),\pi(p_{i+2}))\in E$. Denote the set of shortcuts on $P$ by $C(P)$.
Consider the ideal situation where we are guaranteed that $P$ contains no shortcut (this holds, for example, when $G$ is triangle-free).
In such a case, $(p_1, p_2, p_3)$ should form an induced path and hence the two-hop lemma is always applicable for appropriately chosen $P_1, P_2$.
However, in general, $p_1$ may be a shortcut on $P$, which makes the lemma inapplicable. In this case, one can still use the one-hop lemma, but the resulting probability bound is weaker.
To handle this issue, our first observation is that with high probability, the number of shortcuts on any query path is $O(\log n)$.

\begin{lemma}[Few shortcuts] \label{lem:few-shortcuts}
For every integer $h\geq0$,
\[
 \Pr[\text{some query path has more than $h$ shortcuts}]
 \leq (n+m)2^{-h/2}.
\]
In particular, set $h=\lceil18\log n\rceil$.
With probability at least $1-1/n^7$,
every query path has at most $h$ shortcuts.
\end{lemma}
\begin{proof}
Fix $\pi$ and a query path with $s$ shortcuts. Among the shortcut indices,
at least $s/2$ have the same parity. Their corresponding two-edge segments
are edge-disjoint, although successive segments may share an endpoint.
Choose $r\geq s/2$ such shortcuts. For each subset of these shortcuts,
replace the corresponding two-edge segments by their shortcut edges.
The resulting sequence is still a query path.
Indeed, when $(p_i,p_{i+1},p_{i+2})$ is replaced by $(p_i,p_{i+2})$, the
new query constraint follows from
\[
 p_i<p_{i+1}\leq\inhib(p_{i+2}).
\]
All other query constraints are unchanged. Edge-disjointness guarantees
that these replacements can be made simultaneously. Distinct subsets
omit distinct sets of middle vertices, so they produce $2^r$ distinct
query paths, all ending at $\pi(p_L)$.
Thus, if some path has more than $h$ shortcuts, the total number
$Z(\pi)=\sum_v T(v,\pi)$ of query paths exceeds $2^{h/2}$.
For $h = \lceil 18 \log n\rceil $, $2^{h/2}$ is at least $n^9$.
By \cref{cor:YYI}, $\Pr[Z(\pi) > n^9] \leq \frac{1}{n^7}$.
The probability of having a query path with more than $h$ shortcuts can only be smaller than $\Pr[Z(\pi) > n^9]$.
Hence, the lemma holds.
\end{proof}

Let $A$ be a sufficiently large constant to be determined.
Put
\[
 h=\lceil18\log n\rceil,\qquad
 L=\lceil\sqrt A\log n\rceil,\qquad
 \tau=\lceil18\ln n\rceil,
\]
where $h$ is an upper bound on the number of shortcuts, $L$ is the length of $P$, and $\tau$ is the length of a prefix that will be exposed to satisfy (A3).

Our proof proceeds as follows.
First, we will bound the probability of having a query path
\[
 A\log n<p_1<\cdots<p_L<\beta n,
\]
with at most $h$ shortcuts.
The few-shortcuts lemma will be applied
\emph{after} this bound; we do not condition the random permutation on
the global event that all paths have few shortcuts.
Consider a fixed such position sequence $P$ and a fixed subset
$Q\subseteq\{p_1,\ldots,p_{L-2}\}$ with $|Q|\leq h$, define the event
\[
 \mathcal E(P,Q): P\text{ is a query path and }C(P)=Q.
\]
The set $Q$ is fixed in advance and is not exposed as additional information.

\paragraph{Intervals and protected sets.}
In the following, we upper bound $\Pr[\mathcal{E}(P, Q)]$ using \cref{lem:one-hop,lem:two-hop}.
The proof consists of two components: (1) specifying sets $P_1, P_2$ in the lemmas, and (2) applying a union bound over all possible choices of $(P, Q)$.
We next partition the positions of $\pi$ into geometrically growing intervals and use this partition to assign to each path position $p_k$ a protected set $P_k$ of earlier positions. 
The sizes of these protected sets are chosen so that the $1/\ell = 1/|P_k|$ factors from the bounded-hop lemmas balance -- or cancel out -- the number of possible choices for the path positions.

Let $\alp \in (0, 0.1)$ be a sufficiently small constant determined later.
Divide $\{1, 2, \dots, n\}$ into $I_1, I_2, \dots, I_b$, where
$b=1+\lfloor\log_{1+\alpha}n\rfloor$ and
\[
 I_i=[(1+\alpha)^{i-1},(1+\alpha)^i)\cap\{1,\ldots,n\}
 \quad(1\leq i\leq b).
\]
Let $s_i=|P\cap I_i|$, and write $p_{i,j}$ for the $j$-th position of
$P\cap I_i$. For each $i$ with $s_i>0$, put
\[
 \ell_i=\left\lfloor\frac{|I_{i-1}\setminus P|}{s_i}\right\rfloor.
\]
We say an interval $I_i$ is \emph{occupied} if $s_i > 0$.
For each fixed $i$ with $s_i > 0$, we associate each $p_k \in I_i$ with a protected set $P_k$ constructed as follows.
First, sort $I_{i-1}\setminus P$ and take its first $s_i\ell_i$ positions,
partitioned in order into $s_i$ blocks of size $\ell_i$.
Assign the $j$-th block to $p_{i,j}$ and call it $B_k$ when $p_k=p_{i,j}$.
Thus $B_1<\cdots<B_L$. Define
\[
 P_1=\emptyset,\qquad P_k=B_{k-1}\quad(2\leq k\leq L).
\]

The following lemma establishes that the two-hop lemma is applicable to $(P_k, P_{k+1}, p_k, p_{k+1}, p_{k+2})$ for $k$ satisfying certain conditions; in addition, it establishes simple numerical approximations (such as $\ell_i \approx \frac{|I_{i-1}|}{s_i}$).

\begin{lemma}\label{lem:small-properties}
Fix $\alpha\in(0,1/10)$. For sufficiently large $A$ depending on
$\alpha$, the sets
$P,P_2,\ldots,P_L$ are pairwise disjoint and satisfy:
\begin{itemize}
    \item[(P1)] $P_2<\cdots<P_L$;
    \item[(P2)] $P_k<p_{k-1}$ for every $2\leq k\leq L$;
    \item[(P3)] if $p_k=p_{i,j}$ with $2\leq j<s_i$, then
    $|P_k|=|P_{k+1}|=\ell_i$.
\end{itemize}
\end{lemma}

\begin{proof}
By construction, $B_1<\cdots<B_L$, and the definition $P_k=B_{k-1}$ proves pairwise disjointness and (P1).
For (P2), let $p_{k-1}\in I_r$.
By definition,
\[
    P_k=B_{k-1}\subseteq I_{r-1}<p_{k-1}.
\]

For (P3), suppose $p_k=p_{i,j}$ with $2\leq j<s_i$.
Then $p_{k-1}=p_{i,j-1}$.
The blocks $B_{k-1}$ and $B_k$ are therefore the
$(j-1)$-st and $j$-th blocks allocated from
$I_{i-1}\setminus P$, respectively.
Both have size $\ell_i$, so
\[
    |P_k|=|B_{k-1}|=\ell_i
    =|B_k|=|P_{k+1}|.
\]
\end{proof}

Next, we establish simple numerical approximation (such as $\ell_i \approx \frac{|I_{i-1}|}{s_i}$) that will be useful in the proof.

\begin{claim}
For sufficiently large $A$ depending on $\alp$, every occupied interval $I_i$ satisfies
\begin{equation}\label{eq:rounding}
\begin{gathered}
    \ell_i>10^3, 
    (1-\alpha)\frac{|I_{i-1}|}{s_i}
    \leq \ell_i
    \leq \frac{|I_{i-1}|}{s_i}, 
    \frac{|I_i|}{|I_{i-1}|}\leq 1+2\alpha, \mbox{ and } \tau < \min I_{i-1}
\end{gathered}
\end{equation}
\end{claim}
\begin{proof}
    The proof can be easily checked by elementary calculations, where we use the observation that if $s_i > 0$, then the right endpoint of $I_{i}$ is at least $A\log n$.
\end{proof}

\paragraph{The exposure procedure.}
First expose positions $1,\ldots,\tau$.
Let $\mathcal B$ be the event that
$\Delta(\Galive(\tau+1,n))\leq n/2$.
By \cref{lem:degree-reduction},
\begin{equation}\label{eq:prefix-failure}
 \Pr[\neg\mathcal B]\leq n\exp(-\tau/2)\leq \frac{1}{n^8}.
\end{equation}
Hence, we may assume without loss of generality that $\mathcal{B}$ holds.

Next, expose $p_1$ and $p_2$; the outcome of these two positions can be arbitrary.
The rest of the positions will be exposed via the following iterative procedure.
Start with $k=2$ and
iterate until $k=L$.
We maintain the following invariant:
At the beginning of a step, $p_k$ is the last exposed path position and no protected set $P_r$ with $r\geq k$ has
been exposed.
Let $p_k=p_{i,j}$.
The procedure proceeds as follows.
\begin{itemize}
 \item \textbf{Boundary step.} If $j=1$ or $j\geq s_i-2$, apply the one-hop lemma (\cref{lem:one-hop}). That is, we expose every still unexposed position in $[\max P_k]\setminus P_k$, then expose $P_k$ and $p_{k+1}$.
 After this, we test whether the event in \cref{lem:one-hop} holds for
 $(P_k,p_k,p_{k+1})$, and set $k\gets k+1$. (If the event does not hold, then $P$ does not form a query path. More details will be given later.)
 
 \item \textbf{Interior step without a shortcut.} Otherwise,
 $2\leq j\leq s_i-3$. If $p_k\notin Q$, expose every still unexposed
 position in $[\max P_{k+1}]\setminus(P_k\cup P_{k+1})$.
 Apply the two-hop lemma (\cref{lem:two-hop}) to
 $(P_k,P_{k+1},p_k,p_{k+1},p_{k+2})$, which additionally exposes $P_k, P_{k+1}, p_{k+1}, p_{k+2}$. Set $k\gets k+2$.
 \item \textbf{Interior step with a shortcut.} If instead $p_k\in Q$,
 replace this application of two-hop lemma by two successive applications of the one-hop lemma, for
 $(P_k,p_k,p_{k+1})$ and $(P_{k+1},p_{k+1},p_{k+2})$, with their
 respective prefix exposures. Then set $k\gets k+2$.
\end{itemize}
Note that the procedure is defined based only on $P, Q$ and is independent of the outcome of each exposure step.
Each application of \cref{lem:one-hop,lem:two-hop} is called a \emph{test}.
The test is \emph{successful} if the event in the corresponding lemma holds.

\paragraph{Applicability and necessity of the tests.}
Properties (P1)--(P2) show that filling the prefix before any one-hop
step cannot expose a future protected set or a future path position.
For an interior step, (P1)--(P3) give
\[
 P_k<P_{k+1}<p_k<p_{k+1}<p_{k+2},
 \qquad |P_k|=|P_{k+1}|=\ell_i>10^3.
\]
Thus, filling the prefix for that step also preserves the invariant,
and all exposure assumptions of the two-hop lemma hold.
Note that condition (A3) holds due to the assumption that $\mathcal B$ holds.

If $P$ is a query path, then $p_k$ is alive at time $p_{k-1}$ (for $k \geq 2$). Consequently, when the one-hop lemma is applied to expose some $(P_k, p_k, p_{k+1})$, it is necessary that the event in the lemma holds (i.e., $p_k$ stays alive during the exposure of $P_k$ and $(p_k, p_{k+1}) \in E$).

Consider now an application of the two-hop lemma during the procedure.
Since $p_k \notin Q$, it is necessary that $(p_k, p_{k+1}, p_{k+2})$ forms an induced path.
Hence, $\mathcal E(P,Q)$ implies success of every test in the procedure.

\begin{claim}\label{clm:prob}
Conditioned on any prefix satisfying $\mathcal B$,
\begin{equation}\label{eq:fixed-PQ}
 \Pr[\mathcal E(P,Q)\mid\pi([\tau])]
 \leq (1-\epsilon)^{L/2-1-2b-h}(1-\beta)^{-L}
       \prod_{k=2}^{L-1}\frac{1}{e|P_k|}.
\end{equation}
\end{claim}
\begin{proof}
The procedure covers $L-2$ edges, namely those from $p_k$ to $p_{k+1}$
for $2\leq k\leq L-1$. There are at most four boundary one-hop
steps in each occupied interval. Each interior shortcut step contributes
two further one-hop steps, and there are at most $|Q|\leq h$ such steps.
Thus, if $a$ is the number of one-hop tests and $t$ the number of two-hop
tests, then
\[
 a+2t=L-2 \qquad\mbox{and}\qquad a\leq4b+2h
\]
and hence $t\geq L/2-1-2b-h$.
Each $P_k$, $2\leq k\leq L-1$, contributes exactly one factor
$1/(e|P_k|)$ when it is exposed by either the one-hop or two-hop step.
The conditional probability bounds in the hop lemmas hold regardless of the outcome of all previous exposures.
Therefore, repeated conditioning multiplies the bounds, giving
$(1-\epsilon)^t(1-\beta)^{-a}\prod_{k=2}^{L-1}(e|P_k|)^{-1}$.
To obtain \eqref{eq:fixed-PQ}, we use $a\leq L$ and the lower bound on $t$ to obtain.
\end{proof}

\begin{claim}\label{clm:prob-2}
\begin{equation}\label{eq:block-product}
 \prod_{k=2}^{L-1}\frac{1}{e|P_k|}
 \leq (en)^2(1-\alpha)^{-L}
 \prod_{i:s_i>0}\left(\frac{s_i}{e|I_{i-1}|}\right)^{s_i}.
\end{equation}
\end{claim}
\begin{proof}
The product in the left hand side is over $B_1,\ldots,B_{L-2}$, and omits exactly
$B_{L-1},B_L$. The product over all $L$ blocks is
$\prod_{i:s_i>0}(\frac{1}{e\ell_i})^{s_i}$, since each occupied $I_i$ contains $s_i$ blocks of length $\ell_i$. Restoring the two omitted factors
costs at most $(en)^2$, since every block has size at most $n$.
By \eqref{eq:rounding}, $\ell_i \geq (1-\alp)\frac{|I_{i-1}|}{s_i}$. Hence,
\[
\prod_{k=2}^{L-1}\frac{1}{e|P_k|}
 \leq (en)^2\prod_{i:s_i > 0}(\frac{1}{e\ell_i})^{s_i}
 \leq (en)^2(1-\alpha)^{-L}
 \prod_{i:s_i>0}\left(\frac{s_i}{e|I_{i-1}|}\right)^{s_i}.
\]
This completes the proof.
\end{proof}

Denote by $U_{L, h, \alp, \beta, \eps}$ the term
\[
    (1-\epsilon)^{L/2-1-2b-h}(1-\beta)^{-L}
   (en)^2(1-\alpha)^{-L}.
\]
We omit the subscript when the context is clear.
Combining \cref{clm:prob} and \cref{clm:prob-2} gives the following.

\begin{corollary} \label{corollary:prob}
\[
\Pr[\mathcal E(P,Q)\mid\pi([\tau])]
\leq U_{L,h,\alp,\beta,\eps} \cdot 
 \prod_{i:s_i>0}\left(\frac{s_i}{e|I_{i-1}|}\right)^{s_i},
\]
where $s_i$ denotes $|I_i \cap P|$.
\end{corollary}

As we will show later, $U = (1 - \eps)^{\Theta(L)}$ for sufficiently small $\alp, \beta$ depending on $\eps$ and sufficiently large $A$ depending on $\alp$.

\paragraph{Final union bound.}


We obtain the final probability bound by applying a union bound over all possible $(P, Q)$.
Call the sequence $(s_i)_{1\leq i\leq b}$ the \emph{occupancy vector}.
First, note that the probability upper bound in \cref{corollary:prob} is uniform for a fixed occupancy vector.
We therefore consider a fixed occupancy vector $(s_1,\ldots,s_b)$ and then sum the probability over all $(P, Q)$ satisfying $|P \cap I_i| = s_i, 1 \leq i \leq b$.
The number of such $P$ is at most 
\[
    \prod_{i:s_i>0}\binom{|I_i|}{s_i} \leq \prod_{i:s_i>0}\left( \frac{e|I_i|}{s_i} \right)^{s_i},
\]
which is the number of ways to choose $s_i$ positions from each $I_i$.
The number of choices of $Q$ for each $P$ is at most $\sum_{r=0}^h\binom{L}{r} \leq (h+1) (\frac{eL}{h})^h$, as $Q$ is a subset of $P$ of size at most $h$.
Combining this with \cref{corollary:prob}, the sum of the probabilities for this fixed $(s_i)_{1\leq i \leq b}$ is at most
\begin{align*}
    & U \cdot \prod_{i:s_i>0}\left(\frac{s_i}{e|I_{i-1}|}\right)^{s_i} \cdot \prod_{i:s_i>0}\left( \frac{e|I_i|}{s_i} \right)^{s_i} \cdot (h+1) \left(\frac{eL}{h}\right)^h \\
    \leq~&U \cdot (1 + 2\alp)^L \cdot (h+1) \cdot \left(\frac{eL}{h}\right)^h, & (L \geq 2h)
\end{align*}
where the factors $s_i/(e|I_{i-1}|)$ from \cref{corollary:prob} cancel the factors $e|I_i|/s_i$ from counting positions, up to
$\prod_i(|I_i|/|I_{i-1}|)^{s_i}\leq(1+2\alpha)^L$.

There are at most 
\[
    \binom{L+b-1}{b-1} \leq \left( \frac{e(L + b - 1)}{b - 1}\right)^{b-1}.
\]
occupancy vectors. Hence the
conditional probability of any such path with at most $h$ shortcuts is
at most
\begin{align}
    & U \cdot (1 + 2\alp)^L \cdot (h+1) \cdot \left(\frac{eL}{h}\right)^h \cdot \left( \frac{e(L + b - 1)}{b - 1}\right)^{b-1} \label{eqn:final}
\end{align}

We now show that \cref{eqn:final} is $1/\poly(n)$ for appropriately chosen $A, \alp, \beta$.
Recall that
\begin{equation*}
    b = \log_{1+\alp} n, \quad h = 18 \log n, \quad L = \sqrt{A} \log n, \quad U = (1-\epsilon)^{L/2-1-2b-h}(1-\beta)^{-L}
   (en)^2(1-\alpha)^{-L},
\end{equation*}
where $\eps$ is a constant determined by the proof of \cref{lem:two-hop}, and $A, \alp, \beta$ are parameters that are so far undetermined.
For $\alp,\beta$ sufficiently small depending on $\eps$, the terms $(1+2\alp), (1-\beta)^{-1}$, and $(1-\alp)^{-1}$ are all upper bounded by $(1-\eps)^{-c}$ for some arbitrarily small positive constant $c$.
After fixing $\alp$, the value of $b$ is determined as $O_\alp(\log n)$.
We then choose $A$ to be sufficiently large depending on $\alp,\beta, \eps$ such that 
\begin{equation} \label{eqn:final-1}
    U \cdot (1+2\alp)^{L} \leq (1-\eps)^{L/4}.
\end{equation}
To upper bound the remaining terms, note that
\begin{equation}\label{eqn:final-2}
    (h+1) \cdot \left(\frac{eL}{h}\right)^h \cdot \left( \frac{e(L + b - 1)}{b - 1}\right)^{b-1} \leq \exp(O_{\alp}(\log A \cdot \log n)).
\end{equation}
Since $(1-\eps)^L = \exp(-\Theta(\eps\sqrt{A} \log n))$, \cref{eqn:final-2} is at most $(1-\eps)^{-L/8}$ for sufficiently large $A$ depending on $\eps$ and $\alp$.
Combining this with \cref{eqn:final-1} shows that \cref{eqn:final} is upper bounded by $(1-\eps)^{L/8}$, which is $1/n^7$ for sufficiently large $A$.

If a query path of length $> L$ exists, then its first $L$ vertices form a query path of length exactly $L$. 
Hence, our analysis shows that 
\begin{equation} \label{eqn:general-case}
    \Pr[\text{there is a query path of length $\geq L$ with at most $h$ shortcuts} \mid \mathcal{B} ] \leq 1/n^7.
\end{equation}
To obtain the final probability bound, we union bound over the following three events:
\begin{itemize}
    \item $\neg \mathcal{B}$,
    \item $\Pr[\text{there is a query path of length $\geq L$ with at most $h$ shortcuts} \wedge \mathcal{B}]$, and
    \item $\Pr[\text{there is a query path of length $\geq L$ with more than $h$ shortcuts}]$.
\end{itemize}
Each event happens with probability at most $1/n^7$.
Consequently, we obtain the following.



\begin{lemma}\label{lem:query-path-limited}
There exist absolute constants $A, \beta$ such that, for all
sufficiently large $n$, with probability at least $1-\frac{1}{n^6}$, every query path whose positions lie strictly between
$A\log n$ and $\beta n$ has fewer than $\lceil\sqrt A\log n\rceil$
vertices.
\end{lemma}

\subsection{Concluding the proof} \label{sec:assumptions}
\begin{proof}[Proof of \cref{thm:main}]
Use the fixed absolute constants from \cref{lem:query-path-limited}.
Let $t=\lceil\beta n\rceil-1$ and
$H=\Galive(t+1,n)$. The longest query path contained in positions
$1,\ldots,t$ has at most
\[
 \lfloor A\log n\rfloor+\lceil\sqrt A\log n\rceil=O(\log n)
\]
vertices, except with probability $\frac{1}{n^6}$.
Indeed, remove its initial positions at most $A\log n$ and apply the
lemma to the remaining consecutive query subpath.

By the degree reduction bound, with probability at least
$1-\frac{1}{n^6}$,
\[
 \Delta(H)\leq\frac{8n\ln n}{t}=O(\log n).
\]
Condition on any particular prefix $\pi(1),\ldots,\pi(t)$ satisfying
this degree bound. The vertex set and edges of $H$ are now fixed,
and its vertices inherit a uniformly random relative order from the
remaining permutation. The increasing-path bound in
\cref{lem:increasing-path} implies that $H$ has no increasing path of more than
$O(\log n)$ vertices except with probability $\frac{1}{n^6}$.
Every query path is increasing. Consequently, by
\cref{lem:query-depth-division} with $l=1$ and $r=n$,
\[
 \sigma(G)\leq\sigma(G(1,t))+\sigma(H)+1=O(\log n).
\]
The total failure probability is at most
\[
 \frac{1}{n^6}+\frac{1}{n^6}+\frac{1}{n^6}
 \leq\frac{3}{n^6} \leq \frac{1}{n^7}.
\]
This completes the proof.

\end{proof}

\subsection{Proof of \cref{lem:one-hop}: the one-hop lemma} \label{sec:one-hop}

The intuition of this proof is as follows.
Suppose that the number of neighbors of $\pi(p_1)$ is $D$ for some $D > 0$.
Then, the probability that $\pi(p_1)$ stays alive during the exposure of $P_1$ is $\approx (1 - \frac{D}{n})^{\ell} \leq \exp(-\frac{D\ell}{n})$.
On the other hand, the probability that $\pi(p_2)$ is a neighbor of $\pi(p_1)$ is $\approx \frac{D}{n}$.
The maximum value of the product is $\frac{1}{e\ell}$, attained when $D \approx \frac{n}{\ell}$.
The challenge is mainly to formalize this intuition, as the number of neighbors of $\pi(p_1)$ is changing when $P_1$ is exposed.

We first present several additional definitions.

\begin{definition}[Active vertex, active degree]
A vertex is called \emph{active} at time $t$ if it is unexposed and alive at time $t$.
The \emph{active degree} of a vertex $v$ at time $t$ is the number of its active neighbors.
\end{definition}

\paragraph{The sequential greedy algorithm.}
We consider the following simulation of the sequential greedy algorithm when the positions in $S$ are already exposed.
Initially, at time 1, all vertices are alive, and only the positions in $S$ are exposed.
At time $t$, position $t$ is exposed (if it is not yet exposed).
Then, $\pi(t)$ is processed as follows.
If $\pi(t)$ is alive at time $t$, add $\pi(t)$ to $V^*$ and mark $\pi(t)$ and all its neighbors as dead.
If $\pi(t)$ is already dead at time $t$, the algorithm takes no action.

\paragraph{The auxiliary algorithm.}
We will also work with an \emph{auxiliary algorithm} $\calA$.
Intuitively, $\calA$ represents the behavior of the sequential greedy algorithm conditioned on $\pi(p_1)$ staying alive; the exact definition is slightly different for technical reasons.

The definition of $\calA$ is as follows.
Let $v = \pi(p_1)$ and $t_1 = \max P_1$.
Define $\Omega$ as a set of vertex sequences for $P_1$ such that
\[
    \Omega = \{\omega:P_1 \rightarrow V \mid \text{$v$ is alive at time $t_1 + 1$ if $\pi(P_1) = \omega$} \}.
\]
For $1 \leq i \leq \ell+1$, let $N_1^i$ denote the set of active neighbors of $v$ after the first $i - 1$ positions in $P_1$ are exposed.
For $\omega \in \Omega$, let $d_i(\omega)$ denote $|N_1^i|$ conditioned on $\pi(P_1) = \omega$.
The auxiliary algorithm works exactly as the sequential greedy algorithm with an exception: When $\calA$ exposes a position $P_1(t)$, it selects $\pi(P_1(t))$ uniformly at random from $V \setminus (\pi(1, P_1(t)-1) \cup \pi(S) \cup N_1^t)$ instead of from $V \setminus (\pi(1, P_1(t)-1) \cup \pi(S))$.
In other words, $\calA$ always selects a vertex that will keep $v$ alive (if $v$ is currently alive).
Due to the assumption (A3), the degree of $v$ is at most $0.5n$.
Hence, the set $V \setminus (\pi(1, P_1(t)-1) \cup \pi(S) \cup N_1^t)$ is never empty, as it contains at least $n - \beta n - 0.5n \geq 0.4n$ vertices.

For $\omega \in \Omega$, we denote by $\Pr_{\calA}[\omega \mid \pi(S)]$ the probability that $\calA$ selects $\omega$ for $P_1$.
Note that
\[
    \Pr_{\calA}[\omega \mid \pi(S)] = \prod_{i=1}^{\ell} \frac{1}{n - |S| - (i-1) - d_i(\omega)}.
\]
On the other hand,
\begin{align} 
    \Pr_{P_1}[\pi(P_1) = \omega \mid \pi(S)] &= \prod_{i=1}^{\ell} \frac{1}{n - |S| - (i-1)} \notag \\
    &= \Pr_{\calA}[\omega \mid \pi(S)] \cdot  \prod_{i=1}^{\ell} \paren{1 - \frac{d_i(\omega)}{n-|S|-(i-1)}} \notag \\
    & \leq \Pr_{\calA}[\omega \mid \pi(S)] \cdot \exp\paren{\sum_{i=1}^{\ell}\frac{-d_i(\omega)}{n}}, & \label{eqn:auxiliary}
\end{align}
where the last inequality is due to $1 - x \leq \exp(-x)$.
\cref{eqn:auxiliary} relates $\Pr_\calA[\omega \mid \pi(S)]$ with $\Pr_{P_1}[\pi(P_1) = \omega \mid \pi(S)]$, showing that they are different by a term related to the active degrees of $v$ under $\omega$.

\paragraph{Probability upper bound.}
Recall that the goal is to upper bound the probability of the event
\[
    \calF \defeq \text{$(\pi(p_1), \pi(p_2)) \in E  \text{ and $p_1$ and $p_2$ are alive at time $t_1 + 1$}$}
\]
while exposing $P_1 \cup \{p_2\}$.
Without loss of generality, we assume that $v$ is alive at time $\min P_1$; otherwise, $\Pr[\calF] = 0$ and the lemma is trivially true.

Consider the following process: First, expose $P_1$ by running the sequential greedy algorithm until time $t_1 + 1$. Then, expose $p_2$.
Two events are necessary for $\calF$.
\begin{itemize}
    \item[] $\calF_1$: $v$ stays alive when $P_1$ is exposed, and
    \item[] $\calF_2$: $\pi(p_2)$ is an active neighbor of $v$ when $p_2$ is exposed.
\end{itemize}

Due to $\calF_1$, the vertex sequence for $\pi(P_1)$ must be selected from $\Omega$.
Conditioned on $\pi(P_1) = \omega$ for some $\omega \in \Omega$, when $p_2$ is exposed, the probability of $\calF_2$ is $\frac{d_{\ell+1}(\omega)}{n - |S| - \ell} \leq \frac{d_{\ell+1}(\omega)}{(1-\beta)n}$.
Therefore,
\begin{align*}
    \Pr_{P_1 \cup \{p_2\}}[\calF \mid \pi(S)] &\leq \sum_{\omega \in \Omega} \Pr_{P_1} [\pi(P_1) = \omega \mid \pi(S)] \cdot \frac{d_{\ell+1}(\omega)}{n - |S| - \ell} \\
    & \leq \sum_{\omega \in \Omega} \Pr_{\calA} [\omega \mid \pi(S)] \cdot \exp\paren{\sum_{i=1}^{\ell}\frac{-d_i(\omega)}{n}} \cdot \frac{d_{\ell+1}(\omega)}{(1-\beta)n},
\end{align*}
where the last inequality is due to \cref{eqn:auxiliary}.
Since $(d_i(\omega))_{1 \leq i \leq \ell+1}$ is a non-increasing sequence, the last term maximizes when $d_1(\omega) = d_2(\omega) = \dots = d_\ell(\omega) = D$ for some positive integer $D$.
By differentiation, the maximum is attained when $D = \frac{n}{\ell}$, at which we have
\[
\exp\paren{\sum_{i=0}^{\ell-1}\frac{-D}{n}} \cdot \frac{D}{(1-\beta)n} \leq \frac{1}{e\ell(1-\beta)}.
\]
This completes the proof.

\subsection{Proof of \cref{lem:two-hop}: the two-hop lemma} \label{sec:two-hop}

Let $t_1 = \max P_1$ and $t_2 = \max P_2$.
Consider the following two-phase process.
\begin{itemize}
    \item \textbf{Phase 1:} Run the greedy algorithm until time $t_1 + 1$ to expose $P_1$. Then, expose $p_2$.
    \item \textbf{Phase 2:} Run the greedy algorithm until time $t_2 + 1$ to expose $P_2$. Then, expose $p_3$.
\end{itemize}

Let $v = \pi(p_1)$.
Let $N_1$ and $N_2$ be, respectively, the vertices of distance exactly 1 and 2 from $v$.
Define $\Omega$ and $\calA$ as in the proof of \cref{lem:one-hop} (see \cref{sec:one-hop}).
As in \cref{lem:one-hop}, define $N_1^i$ as the set of active vertices in $N_1$ after the first $i-1$ positions of $P_1$ are exposed.
For $\omega \in \Omega$, let $N_1^i(\omega)$ be $N_1^i$ conditioned on $\pi(P_1) = \omega$.
Similarly, define $N_2^i$ as the set of active vertices in $N_2$ after the first $i-1$ positions of $P_1$ are exposed.

We say Phase 1 is \emph{successful} if $\pi(P_1) \in \Omega$ and $\pi(p_2) \in N_1^{\ell+1}$.
Conditioned on $\pi(p_2) = u$ for some $u \in V$ and $\pi(P_1) = \omega$ for $\omega \in \Omega$, we say Phase 2 is \emph{successful} if $u$ stays alive after $P_2$ is exposed and $\pi(p_3)$ is an active neighbor of $u$ in $N_2$ when $p_3$ is exposed.
Note that the event $\calG$ holds if and only if both phases are successful.


Let $n' = (1-\beta) n$.
Denote by $H(\omega, u)$ the success probability of Phase 2 conditioned on $\pi(P_1) = \omega$ and $\pi(p_2) = u$.
To succeed in both phases, it is necessary that $\pi(P_1) \in \Omega$, $\pi(p_2) \in N_1^{\ell+1}$, and under these conditions we succeed in Phase 2.
Hence, 
\begin{align}
    \Pr[\calG] &= \sum_{\omega \in \Omega, u \in N_1^{\ell+1}(\omega)} \Pr_{P_1}[\pi(P_1) = \omega \mid \pi(S)] \cdot \Pr_{p_2}[\pi(p_2) = u \mid \pi(S), \pi(P_1) = \omega] \cdot  H(\omega, u) \notag  \\ 
    &\leq \sum_{\omega \in \Omega, u \in N_1^{\ell+1}(\omega)} \Pr_{\calA}[\omega \mid \pi(S)] \cdot \exp\paren{\sum_{i=1}^{\ell}\frac{-d_i(\omega)}{n}} \cdot \frac{1}{n'} \cdot H(\omega, u) \notag \\
    &= \sum_{\omega \in \Omega} \Pr_\calA [\omega \mid \pi(S)] \cdot \exp\paren{\sum_{i=1}^{\ell} \frac{-d_i(\omega)}{n}} \cdot \sum_{u \in N_1^{\ell+1}(\omega)}\paren{\frac{1}{n'} \cdot H(\omega, u)}, \label{eqn:two-hop-1} 
\end{align}
where the first inequality is due to \cref{eqn:auxiliary} and $\Pr_{p_2}[\pi(p_2) = u \mid \pi(S), \pi(P_1) = \omega] \leq \frac{1}{n'}$.

By the proof of \cref{lem:one-hop}, we have that $H(\omega, u) \leq \frac{1}{e\ell(1-\beta)}$ for any $\omega \in \Omega$, $u \in V$.
However, the probability of $\frac{1}{e\ell(1-\beta)}$ is only achieved if the active degree of $u$ is $\frac{n}{\ell}$.
We say a vertex $v \in N_1^i$ is a \emph{candidate in $N_1^i$} if it has at least $0.9\frac{n}{\ell}$ active neighbors in $N_2^i$.

\begin{claim} \label{clm:candidate}
    Consider an $\omega \in \Omega$ and $u \in V$.
    Suppose that $u$ is not a candidate in $N_1^{\ell+1}(\omega)$ conditioned on $\pi(P_1) = \omega$.
    Then, there exists a constant $\eps'$ such that
    \[
        H(\omega, u) \leq \frac{1-\eps'}{e\ell}.
    \]
\end{claim}
\begin{proof}
The proof of this claim is analogous to that of the one-hop lemma (\cref{lem:one-hop}).
The required modifications are as follows.
For Phase 2 to be successful, $u$ has to stay alive when $P_2$ is exposed, and $\pi(p_3)$ has to be an active neighbor of $u$ in $N_2$.
Compared to the one-hop lemma, the only difference is the additional requirement that $\pi(p_3) \in N_2$.

By repeating the argument in the proof of \cref{lem:one-hop}, the worst case of Phase 2 happens when the active degree of $u$ is $D = \frac{n}{\ell}$ when $P_2$ is exposed, and all these edges are between $u$ and $N_2$.
Since $u$ is not a candidate in $N_1^{\ell+1}$ when $\pi(P_1) = \omega$, the number of active neighbors of $u$ in $N_2$ is a constant factor smaller than $D$.
It is not hard to verify that the worst-case probability is a constant factor smaller under this condition.
\end{proof}

Denote by $f(\omega)$ the number of candidates in $N_1^{\ell+1}(\omega)$.
Let $\delta = 0.1$.
The assumption that $\ell \geq 10^3$ ensures $\lfloor \delta \ell\rfloor \geq 0.99\delta \ell$.
Consider the following three types of vertex sequences in $\Omega$.
\begin{itemize}
    \item $\Omega_1 = $ the set of vertex sequences $\omega \in \Omega$ such that $d_{\lfloor \delta\ell \rfloor}(\omega) \geq 1.1\frac{n}{\ell}$ or $d_{\ell+1}(\omega) < 0.9 \frac{n}{\ell}$.
    In this case, we use the simple upper bound $H(\omega, u) \leq \frac{1}{e\ell(1-\beta)}$.
    Hence, for $\omega \in \Omega_1$, the summand in \cref{eqn:two-hop-1} is at most
    \begin{equation*}
        \Pr_\calA [\omega \mid \pi(S)] \cdot \exp\paren{\sum_{i=1}^{\ell} \frac{-d_i(\omega)}{n}} \cdot \frac{d_{\ell+1}(\omega)}{n'} \cdot \frac{1}{e\ell(1-\beta)}.
    \end{equation*}
    As shown in \cref{lem:one-hop}, the maximum of $\exp\paren{\sum_{i=1}^{\ell} \frac{-d_i(\omega)}{n}} \cdot \frac{d_{\ell+1}(\omega)}{n'}$ is $\frac{1}{e\ell(1-\beta)}$, and it is attained when $d_1(\omega) = d_2(\omega) = \dots = d_{\ell+1}(\omega) = \frac{n}{\ell}$.
    For $\omega \in \Omega_1$, $\{d_i(\omega)\}_{1\leq i \leq \ell+1}$ is sufficiently away from the optimizer, and thus the worst-case probability is improved by a constant factor $(1-\eps_1)$.
    
    \item $\Omega_2 =$ the set of vertex sequences $\omega \in (\Omega \setminus \Omega_1)$ such that $f(\omega) \leq 0.8\frac{n}{\ell}$.
    For $\omega \in \Omega_2$, it is likely that $\pi(p_2)$ is a non-candidate.
    By \cref{clm:candidate}, when $\pi(p_2)$ is a non-candidate, $H(\omega, u)$ is improved by a constant factor. Hence, we bound the term
    \begin{align}
        \sum_{u \in N_1^{\ell+1}(\omega)}\paren{\frac{1}{n'} \cdot H(\omega, u)}  \leq \frac{f(\omega)}{n'} \cdot \frac{1}{e\ell(1-\beta)} + \frac{d_{\ell+1}(\omega) - f(\omega)}{n'} \cdot \frac{1 - \eps'}{e\ell} \leq \frac{d_{\ell+1}(\omega)}{n'} \cdot \frac{1-\eps_2}{e\ell}
    \end{align}
    for some constant $\eps_2$;
    note that the second inequality holds because $f(\omega) \leq 0.8\frac{n}{\ell}$ and $d_{\ell+1}(\omega) > 0.9 \frac{n}{\ell}$.
    Hence, we can upper bound the summand in \cref{eqn:two-hop-1} as
    \[
        \Pr_\calA [\omega \mid \pi(S)] \cdot \exp\paren{\sum_{i=1}^{\ell} \frac{-d_i(\omega)}{n}} \cdot \frac{d_{\ell+1}(\omega)}{n'} \cdot \frac{1-\eps_2}{e\ell} \leq \Pr_\calA [\omega \mid \pi(S)] \cdot \frac{1-\eps_2}{(e\ell)^2(1-\beta)},
    \]
    yielding a constant factor improvement over the $\frac{1}{(e\ell)^2}$ upper bound for $\beta$ small enough.
    
    \item $\Omega_3 = $ the set $\Omega \setminus \Omega_1 \setminus \Omega_2$. We claim that $\sum_{\omega \in \Omega_3} \Pr_{\calA}[\omega \mid \pi(S)] \leq 1 - \eps_3$ for some constant $\eps_3 > 0$.
    Consider a fixed vertex $u \in N_1^1$.
    Let $\calF_i$ be the event that $u$ is a candidate in $N_{1}^i$.
    Conditioned on $\calF_{i-1}$ and $\pi(S)$, $\calF_i$ holds only if $\calA$ does not select an active neighbor of $u$ in $N_2^{i-1}$.
    Due to $\calF_{i-1}$, $u$ is a candidate in $N_1^{i-1}$ and hence there are at least $0.9 \frac{n}{\ell}$ such active neighbors.
    Hence, for $1 < i \leq \ell+1$, we have
    \begin{equation} \label{eqn:kill-candidate}
        \Pr_\calA [\calF_i \mid \calF_{i-1}, \pi(S)] \leq 1 - \frac{0.9n/\ell}{n} \leq \exp(-\frac{0.9}{\ell}).
    \end{equation}
    Since $\omega \notin \Omega_1$, $d_{\lfloor \delta \ell \rfloor}(\omega) \leq 1.1 \frac{n}{\ell}$ and thus there are at most $1.1 \frac{n}{\ell}$ candidates in $N_1^{\lfloor \delta \ell \rfloor}$.
    By repeatedly applying \cref{eqn:kill-candidate}, a candidate $u$ in $N_1^{\lfloor \delta \ell \rfloor}$ remains a candidate in $N_1^{\ell+1}$ with probability at most
    \[
        \exp(-\frac{0.9}{\ell} \cdot (1  - \delta)\ell) \leq \exp(-0.81) \leq 0.5.
    \]
    Hence, $\E_\calA[f(\omega) \mid d_{\lfloor\delta\ell\rfloor}(\omega) < 1.1 \frac{n}{\ell}, \pi(S)] \leq 0.55 \frac{n}{\ell}$.
    By Markov's inequality, $\Pr_\calA[ f(\omega) > 0.8 \frac{n}{\ell}\mid d_{\lfloor\delta\ell\rfloor}(\omega) < 1.1 \frac{n}{\ell}, \pi(S)] \leq 1 - \eps_3$ for some constant $\eps_3$.
    Each $\omega \in \Omega_3$ satisfies $d_{\lfloor \delta\ell \rfloor}(\omega) < 1.1\frac{n}{\ell}$ and $f(\omega) > 0.8\frac{n}{\ell}$.
    Hence, $\sum_{\omega \in \Omega_3} \Pr_\calA[\omega \mid \pi(S)] \leq 1- \eps_3$.
\end{itemize}

Combining the analysis of the three cases, \cref{eqn:two-hop-1} is upper bounded by
\begin{align*}
    \sum_{\omega \in \Omega_1 \cup \Omega_2} \Pr_\calA[\omega \mid \pi(S)] \cdot \frac{1 - \min(\eps_1, \eps_2)}{(e\ell)^2(1-\beta)^2} + \sum_{\omega \in \Omega_3} \Pr_{\calA} [\omega \mid \pi(S)] \cdot \frac{1}{(e\ell)^2(1-\beta)^2} \leq \frac{1-\eps}{(e\ell)^2}
\end{align*}
for some constant $\eps$ and sufficiently small $\beta$.
This completes the proof.

    
\section*{Acknowledgments}
\paragraph{AI disclosure.} GPT-5.4 Pro, GPT-5.6 Sol, and GPT-6 Astra were used in this research project.
The project started from a series of discussions with GPT-5.4 Pro about extending \cite{fischer2019tight}'s work to analyzing query depth.
The output was not used in the rest of the project.
Later, the authors found a proof of $O(\log n)$ query depth in triangle-free graphs, which is essentially the proof in this paper when the set of shortcuts is empty.
This idea and several different attempts to extend it were presented to GPT-5.6 Sol.
The tool outputs a key observation, which is later adapted to \cref{cor:YYI}, showing that only shortcut sets of size $O(\log n)$ are relevant.
This key observation is crucial to extend the result on triangle-free graphs to general graphs.

AI tools, including GPT-6 Astra, were also used to assist in proofreading and creating figures from given drawings. The authors are fully responsible for the integrity, accuracy, and originality of the submission.

\bibliographystyle{alpha}
\bibliography{reference}

@inproceedings{blelloch2012greedy,
author = {Blelloch, Guy E. and Fineman, Jeremy T. and Shun, Julian},
title = {Greedy sequential maximal independent set and matching are parallel on average},
year = {2012},
isbn = {9781450312134},
publisher = {Association for Computing Machinery},
address = {New York, NY, USA},
url = {https://doi.org/10.1145/2312005.2312058},
doi = {10.1145/2312005.2312058},
booktitle = {Proceedings of the Twenty-Fourth Annual ACM Symposium on Parallelism in Algorithms and Architectures},
pages = {308–317},
numpages = {10},
location = {Pittsburgh, Pennsylvania, USA},
series = {SPAA '12}
}

@article{fischer2019tight,
author = {Fischer, Manuela and Noever, Andreas},
title = {Tight Analysis of Parallel Randomized Greedy MIS},
year = {2019},
issue_date = {January 2020},
publisher = {Association for Computing Machinery},
address = {New York, NY, USA},
volume = {16},
number = {1},
issn = {1549-6325},
url = {https://doi.org/10.1145/3326165},
doi = {10.1145/3326165},
journal = {ACM Trans. Algorithms},
month = dec,
articleno = {6},
numpages = {13}
}

@inproceedings{behnezhad2022almost,
  author       = {Soheil Behnezhad and
                  Moses Charikar and
                  Weiyun Ma and
                  Li{-}Yang Tan},
  title        = {Almost 3-Approximate Correlation Clustering in Constant Rounds},
  booktitle    = {63rd {IEEE} Annual Symposium on Foundations of Computer Science, {FOCS}
                  2022, Denver, CO, USA, October 31 - November 3, 2022},
  pages        = {720--731},
  publisher    = {{IEEE}},
  year         = {2022},
  url          = {https://doi.org/10.1109/FOCS54457.2022.00074},
  doi          = {10.1109/FOCS54457.2022.00074},
  bibsource    = {dblp computer science bibliography, https://dblp.org}
}

@article{behnezhad2021massively,
  title={Massively parallel computation via remote memory access},
  author={Behnezhad, Soheil and Dhulipala, Laxman and Esfandiari, Hossein and {\L}{\k{a}}cki, Jakub and Mirrokni, Vahab and Schudy, Warren},
  journal={ACM Transactions on Parallel Computing},
  volume={8},
  number={3},
  pages={1--25},
  year={2021},
  publisher={ACM New York, NY}
}

@article{karp1985fast,
  title={A fast parallel algorithm for the maximal independent set problem},
  author={Karp, Richard M and Wigderson, Avi},
  journal={Journal of the ACM (JACM)},
  volume={32},
  number={4},
  pages={762--773},
  year={1985},
  publisher={ACM New York, NY, USA}
}

@inproceedings{luby1985simple,
  title={A simple parallel algorithm for the maximal independent set problem},
  author={Luby, Michael},
  booktitle={Proceedings of the seventeenth annual ACM symposium on Theory of computing},
  pages={1--10},
  year={1985}
}

@article{alon1986fast,
  title={A fast and simple randomized parallel algorithm for the maximal independent set problem},
  author={Alon, Noga and Babai, L{\'a}szl{\'o} and Itai, Alon},
  journal={Journal of algorithms},
  volume={7},
  number={4},
  pages={567--583},
  year={1986},
  publisher={Elsevier}
}

@article{calkin1990probabilistic,
  title={Probabilistic analysis of a parallel algorithm for finding maximal independent sets},
  author={Calkin, Neil and Frieze, Alan},
  journal={Random Structures \& Algorithms},
  volume={1},
  number={1},
  pages={39--50},
  year={1990},
  publisher={Wiley Online Library}
}

@inproceedings{panconesi1992improved,
  title={Improved distributed algorithms for coloring and network decomposition problems},
  author={Panconesi, Alessandro and Srinivasan, Aravind},
  booktitle={Proceedings of the twenty-fourth annual ACM symposium on Theory of computing},
  pages={581--592},
  year={1992}
}

@inproceedings{naor1993can,
  title={What can be computed locally?},
  author={Naor, Moni and Stockmeyer, Larry},
  booktitle={Proceedings of the twenty-fifth annual ACM symposium on Theory of computing},
  pages={184--193},
  year={1993}
}

@inproceedings{kuhn2005fast,
  title={Fast deterministic distributed maximal independent set computation on growth-bounded graphs},
  author={Kuhn, Fabian and Moscibroda, Thomas and Nieberg, Tim and Wattenhofer, Roger},
  booktitle={International Symposium on Distributed Computing},
  pages={273--287},
  year={2005},
  organization={Springer}
}

@inproceedings{levi2015local,
  title={Local computation algorithms for graphs of non-constant degrees},
  author={Levi, Reut and Rubinfeld, Ronitt and Yodpinyanee, Anak},
  booktitle={Proceedings of the 27th ACM symposium on Parallelism in Algorithms and Architectures},
  pages={59--61},
  year={2015}
}

@inproceedings{ghaffari2016improved,
  title={An improved distributed algorithm for maximal independent set},
  author={Ghaffari, Mohsen},
  booktitle={Proceedings of the twenty-seventh annual ACM-SIAM symposium on Discrete algorithms},
  pages={270--277},
  year={2016},
  organization={SIAM}
}

@inproceedings{ghaffari2022local,
  title={Local computation of maximal independent set},
  author={Ghaffari, Mohsen},
  booktitle={2022 IEEE 63rd Annual Symposium on Foundations of Computer Science (FOCS)},
  pages={438--449},
  year={2022},
  organization={IEEE}
}

@inproceedings{khoury2026breaking,
  title={Breaking barriers for distributed mis by faster degree reduction},
  author={Khoury, Seri and Schild, Aaron},
  booktitle={Proceedings of the 58th Annual ACM Symposium on Theory of Computing},
  pages={1037--1048},
  year={2026}
}

@inproceedings{yoshida2009improved,
  title={An improved constant-time approximation algorithm for maximum matchings},
  author={Yoshida, Yuichi and Yamamoto, Masaki and Ito, Hiro},
  booktitle={Proceedings of the 41st Annual ACM Symposium on Theory of Computing (STOC 2009)},
  pages={225--234},
  year={2009}
}

@inproceedings{onak2012near,
  title={A near-optimal sublinear-time algorithm for approximating the minimum vertex cover size},
  author={Onak, Krzysztof and Ron, Dana and Rosen, Michal and Rubinfeld, Ronitt},
  booktitle={Proceedings of the 23rd Annual ACM-SIAM Symposium on Discrete Algorithms (SODA 2012)},
  pages={1123--1131},
  year={2012},
  organization={SIAM}
}

@inproceedings{behnezhad2021time,
  title={Time-optimal sublinear algorithms for matching and vertex cover},
  author={Behnezhad, Soheil},
  booktitle={Proceedings of the 62nd Annual IEEE Symposium on Foundations of Computer Science (FOCS 2021)},
  pages={873--884},
  year={2021},
  organization={IEEE}
}

@inproceedings{mahabadi2025sublinear,
  title={Sublinear metric Steiner tree via improved bounds for set cover},
  author={Mahabadi, Sepideh and Roghani, Mohammad and Tarnawski, Jakub and Vakilian, Ali},
  booktitle={Proceedings of the Innovations in Theoretical Computer Science Conference (ITCS 2025)},
  pages={74--1},
  year={2025},
  organization={Schloss Dagstuhl--Leibniz-Zentrum f{\"u}r Informatik}
}

@inproceedings{mahabadi2026sublinear,
  title={Sublinear metric Steiner forest via maximal independent set},
  author={Mahabadi, Sepideh and Roghani, Mohammad and Tarnawski, Jakub and Vakilian, Ali},
  booktitle={Proceedings of the Annual ACM-SIAM Symposium on Discrete Algorithms (SODA 2026)},
  pages={5941--5959},
  year={2026},
  organization={SIAM}
}

@inproceedings{azarmehr2025stochastic,
  title={Stochastic matching via in-n-out local computation algorithms},
  author={Azarmehr, Amir and Behnezhad, Soheil and Ghafari, Alma and Rubinfeld, Ronitt},
  booktitle={Proceedings of the Annual ACM Symposium on Theory of Computing (STOC 2025)},
  pages={1055--1066},
  year={2025}
}

@inproceedings{nguyen2008constant,
  title={Constant-time approximation algorithms via local improvements},
  author={Nguyen, Huy N and Onak, Krzysztof},
  booktitle={Proceedings of the 49th Annual IEEE Symposium on Foundations of Computer Science (FOCS 2008)},
  pages={327--336},
  year={2008},
  organization={IEEE}
}

@article{coppersmith1989parallel,
  title={Parallei graph algorithms that are efficient on average},
  author={Coppersmith, Don and Raghavan, Prabhakar and Tompa, Martin},
  journal={Information and Computation},
  volume={81},
  number={3},
  pages={318--333},
  year={1989},
  publisher={Elsevier}
}

@article{calkin1992expected,
  title={On the expected performance of a parallel algorithm for finding maximal independent subsets of a random graph},
  author={Calkin, Neil J and Frieze, Alan M. and Ku{\v{c}}era, Lǔdek},
  journal={Random Structures \& Algorithms},
  volume={3},
  number={2},
  pages={215--221},
  year={1992},
  publisher={Wiley Online Library}
}

\appendix
\section{Lower bounds} \label{sec:lower-bound}
The following construction supplies the lower bound for \cref{thm:main}.

\begin{lemma} \label{lem:lower-bound}
    For any positive integer $\gamma > 3$ and $n_0 > \gamma^{10}$, there exists a graph $G$ with $|V(G)| \geq n_0$ such that, with probability at least $n^{-\frac{2}{\gamma-1}}$, the query depth is $\Omega(\log_{\gamma} n)$.
\end{lemma}
\begin{proof}
The construction is as follows.
First, construct $k+1$ disjoint vertex sets $S_0, S_1, \dots, S_k$ such that $|S_i| = \gamma^i$ for each $i$ and $|S_k| \in [n_0, \gamma n_0)$.
The total number of vertices is $n \defeq \frac{\gamma^{k+1} - 1}{\gamma - 1}$.
Note that $n_0 \leq n \leq 2\gamma n_0$.
Then, for each $0 \leq i < k$, we add a complete bipartite graph between $S_i$ and $S_{i+1}$.

Let $u_i$ be the vertex with the smallest rank in $S_i$.
Denote by $\calE$ the event that $\rank(u_k) < \rank(u_{k-1}) < \dots < \rank(u_0)$.
Consider an invocation of $\RGMIS$ on $u_0$.
Suppose that $\calE$ holds.
The recursion must explore the path $(u_0, u_1, \dots, u_k)$.
The length of the path is $\Omega(\log_{\gamma} n_0) = \Omega(\log_{\gamma} n)$.
Hence, if $\calE$ holds, the query depth is $\Omega(\log_{\gamma} n)$.
To prove the lemma, it suffices to show that $\Pr[\calE]$ is at least $n^{-\frac{2}{(\gamma-1)}}$.

Denote by $S'_i = \bigcup_{0 \leq j \leq i} S_j$.
Let $\calE_i$ be the event that the minimum-rank vertex among $S'_i$ is in $S_i$.
Note that $\calE = \bigcap_{0\leq i \leq k} \calE_i$.
The probability of $\calE_k$ is
\[
    \frac{|S_k|}{|S'_k|} \geq \frac{\gamma^k}{\gamma^k+\frac{\gamma^k-1}{\gamma-1}} \geq \frac{1}{1+\frac{1}{\gamma-1}} \geq \exp(\frac{-1}{\gamma-1}).
\]
Conditioned on $\calE_k$, the probability of $\calE_{k-1}$ is $\frac{|S_{k-1}|}{|S'_{k-1}|}$, which again is at least $\exp(\frac{-1}{\gamma-1})$.
More generally, the probability of $\calE_i$ conditioned on $\bigcap_{j>i} \calE_{j}$ is $\frac{|S_{i}|}{|S'_{i}|} \geq \exp(\frac{-1}{\gamma-1})$, since $\bigcap_{j>i} \calE_{j}$ does not depend on the internal ordering of $S'_i$.
Hence, $\Pr[\calE] \geq \exp(\frac{-k}{\gamma-1}) = \exp(\frac{-\ceil{\log_{\gamma} n_0}}{\gamma-1}) \geq n^{-\frac{2}{(\gamma-1)}}$.
This completes the proof.
\end{proof}

In \cref{lem:lower-bound}, the constructed graph is triangle-free (in fact, it is bipartite).
Taking $\gamma=4$ gives a query depth of $\Omega(\log n)$ with
probability at least $n^{-2/3}$.
Hence, \cref{thm:main} is tight.
(More precisely, one cannot replace $O(\log n)$ with any function in $o(\log n)$ in this theorem and requires it to still hold with high probability.)

\paragraph{Lower bound on paths.}
The following standard example shows that the general upper bound is
within an $O(\log\log n)$ factor of the query depth even on paths.

\begin{lemma}[Folklore] \label{lem:path-lower-bound}
The query depth of a length-$n$ path is $\Omega(\log n / \log\log n)$ with high probability.
\end{lemma}
\begin{proof}
Consider a sub-path $P' = (v_1, v_2, \dots, v_k)$ of the path.
Observe that if $P'$ is an increasing path, then $(v_1, v_{2}, \dots, v_{k-1})$ must be a query path; in that case, the query depth is at least $k - 1$.

Note that $\Pr[P' \text{ is an increasing path}] = \frac{1}{k!} \geq \frac{1}{2^{k \log k}}$.
Take $k = \left\lfloor\frac{\log n}{2\log\log n}\right\rfloor$.
We have $2^{k \log k} \leq 2^{\log n / 2}$ and thus $\frac{1}{2^{k \log k}} \geq \frac{1}{\sqrt{n}}$.
For sufficiently large $n$, a path of $n$ vertices has more than $10\sqrt{n} \ln n$ disjoint sub-paths of length $k$.
The probability that at least one of them is an increasing path is at least
\[
    1 - (1-1/\sqrt{n})^{10\sqrt{n} \ln n} \geq 1- \exp(-10 \ln n) = 1 - 1/n^{10}.
\]
Hence, the query depth is at least $k-1=\Omega(\log n/\log\log n)$
with high probability.
\end{proof}

Thus, on paths, \cref{thm:main} is tight up to an $O(\log\log n)$ factor.

\end{document}